\documentclass[a4paper,11pt]{article}

\usepackage{amsfonts,amssymb,amsmath,amsthm,colonequals}
\usepackage[mathscr]{euscript}
\usepackage{float}
\usepackage{caption}
\usepackage{cite}
\usepackage{tabularx}
\usepackage[british]{babel}
\usepackage{amssymb}
\usepackage{amsmath}
\usepackage{mathrsfs}

\newcommand\fverb{\setbox\fverbbox=\hbox\bgroup\verb}
\newcommand\fverbdo{\egroup\medskip\noindent
			\fbox{\unhbox\fverbbox}\ }
\newcommand\fverbit{\egroup\item[\fbox{\unhbox\fverbbox}]}
\newbox\fverbbox

\newcommand{\defeq}{\stackrel{\textup{\tiny def}}{=}}

\newcommand{\beqa}{\begin{eqnarray}}
\newcommand{\eeqa}{\end{eqnarray}}
\newcommand{\beq}{\begin{equation}}
\newcommand{\eeq}{\end{equation}}
\DeclareMathOperator{\g}{\mathfrak{g}}
\DeclareMathOperator{\h}{\mathfrak{h}}

\newcommand{\calS}{\mathcal{S}}

\newcommand{\calO}{\mathcal{O}}

\newcommand{\calD}{\mathcal{D}}

\newcommand{\fg}{\mathfrak{g}}\newcommand{\fh}{\mathfrak{h}}
\newcommand{\fm}{\mathfrak{m}}

\newcommand\id{{\bf 1}}

\DeclareMathOperator{\Cas}{\mathbf{Cas}}

\newcommand{\vac}[1]{\ensuremath{\left< \, #1\, \right>}}

\newcommand{\com}[2]{\ensuremath{\left[ #1\, ,\, #2\right]}}

\newcommand{\form}[2]{\ensuremath{\left( #1, #2\right)}}

\newcommand{\tb}[1]{\textbf{#1}}
\newcommand{\A}{{\alpha}}
\newcommand{\B}{{\beta}}
\newcommand{\C}{\gamma}
\newcommand{\D}{{\delta}}
\newcommand{\E}{{\epsilon}}
\newcommand{\F}{{\varphi}}
\newcommand{\G}{{\rho}}
\renewcommand{\H}{{\chi}}
\newcommand{\I}{\mathcal{I}}

\newcommand{\half}{\ensuremath{\frac{1}{2}}}

\newcommand{\dd}{\text{d}}
\newcommand{\fact}[1]{#1!\,}
\newcommand{\grade}[1]{\lvert#1\rvert}
\newcommand{\abs}[1]{\left\lvert#1\right\rvert}
\newcommand{\order}[1]{\ensuremath{\mathcal{O}}(#1)}
\newcommand{\orderepsilon}{\order{\varepsilon}}

\newcommand{\dom}{\mathbb{C}_{\varepsilon}}

\newcommand{\oep}{\orderepsilon}

\newcommand{\widebar}{\bar}

\begin{document}
\thispagestyle{empty}
\setcounter{page}{0}
\begin{flushright}\footnotesize
\texttt{DESY 14-229}\\
\footnotesize
\texttt{NORDITA-2015-140}\\
\vspace{2.5cm}
\end{flushright}
\setcounter{footnote}{0}

\begin{center}
{\LARGE\tb{\mathversion{bold}
Spectra of Sigma Models on \\[2mm]
Semi-Symmetric Spaces}\par}
\vspace{15mm}

{\sc  Alessandra Cagnazzo, Volker Schomerus, Vaclav Tlapak}\\[5mm]

{\it DESY Hamburg, Theory Group, \\
Notkestrasse 85, D--22607 Hamburg, Germany
}\\[5mm]

\texttt{alessandra.cagnazzo@desy.de}\\
\texttt{volker.schomerus@desy.de}\\
\texttt{vaclav.tlapak@desy.de}\\[25mm]

\tb{Abstract}\\[2mm]
\begin{minipage}{1.0\columnwidth}
Sigma models on semi-symmetric spaces provide the central building
block for string theories on $AdS$ backgrounds. Under certain conditions
on the global supersymmetry group they can be made one-loop conformal by
adding an appropriate fermionic Wess-Zumino term. We determine the full
one-loop dilation operator of the theory. It involves an interesting new
XXZ-like interaction term. Eigenvalues of our dilation operator, i.e.\ the
one-loop anomalous dimensions, are computed for a few examples.
\end{minipage}
\end{center}
\newpage
\tableofcontents

\section{Introduction}
Non linear sigma models (NLSM), such as the famous $O(N)$ or $\mathbb{CP}^N$
models, play an important role in high and low energy physics as well as mathematics.
For a long time, research focused on cases in which the target is a symmetric
space, i.e.\ can be written as a quotient $G/H$ of a (super)group G by a
subgroup $H \subset G$ that is invariant under the action of an involution
$\sigma: G \rightarrow G$, i.e.\ by an automorphism of order two. The AdS/CFT
correspondence has brought more general homogenous spaces into the
spotlight, including
$$ PSU(2,2|4)\, \bigl{/} SO(1,4) \times SO(5)  \quad
     \mbox{ and } \quad
   SPO(2,2|6)\, \bigl{/} SO(1,3) \times U(3) \  $$
which describe strings moving on $AdS_5 \times S^5$ and $AdS_4 \times
\mathbb{CP}^3$, respectively. These spaces are examples of so-called
generalized symmetric spaces $G/H$. By definition, the stabilizer
sub(super)group $H$ of a generalized symmetric space is left
invariant by the action of an automorphism $\sigma$ of order $M>2$.
Sigma models on such generalized symmetric spaces are not uniquely
fixed by the target space manifold since the $G$ symmetry leaves us with
some $M-1$ dimensional space of metrics. Additionally, it is possible
to add $\theta$ or Wess-Zumino (WZ) terms.

Since all the known relevant examples, such as the two displayed above,
involve automorphisms or order four, we shall restrict to $M=4$. The
corresponding coset spaces $G/H$ are often referred to as semi-symmetric
and their sigma models appear as a part of the world-sheet action
for strings in homogeneous AdS backgrounds, regardless of whether one works
within the Green-Schwarz \cite{Metsaev:1998it,Bena:2003wd,Babichenko:2009dk,
Sorokin:2011rr,Arutyunov:2008if},
pure spinor \cite{Berkovits:2000fe} or hybrid formalism \cite{Berkovits:1999im,Berkovits:1999zq}. Only the choice of couplings
depends on which formulation of superstring theory is actually being
used. In this paper we are not concerned with the relations between the
different approaches \cite{Berkovits:2001ue,D'Auria:2008ny,Tonin:2010mm,
Oda:2011nc,Aisaka:2012ud,Cagnazzo:2012uq} and simply pick the couplings
in the action such that we recover the NLSM of the hybrid and pure spinor
models. In these cases, the metric conspires with a fermionic WZ term in
order to make the action classically integrable and the one-loop
beta-function vanish \cite{Young:2005jv}.

The aim of this work is to study the spectrum of one-loop anomalous dimensions
for one-loop conformal NLSMs on semi-symmetric superspaces. Our analysis
is valid for all such models, compact and non-compact. In order to fully
control the one-loop spectrum, one also needs to enumerate possible
vertex operators in the free theory. This is a problem of harmonic analysis
which will not be addressed in the present work, see \cite{Candu:2013cga},
however, where the corresponding issue has been solved for compact
supercoset geometries. The analysis of \cite{Candu:2013cga} carries over to
non-compact cosets as long as the stabilizer subgroup $H$ is compact and a
generalization to noncompact $H$ is possible at least on a case-by-case basis.

Our results for the full one-loop dilation operator of  one-loop conformal
NLSMs on semi-symmetric spaces will be given in section \ref{sec:summary}.
For non-derivative fields (zero modes) the one-loop dilation operator is
given by the Bochner Laplacian, just as in the case of symmetric superspaces,
see \cite{Candu:2013cga}. For operators including world-sheet derivatives the
results becomes more interesting. In this case, the one-loop dilation operator
turns out to involve an interesting new XXZ-like interaction term that we
introduce in section \ref{sec:summary} and derive in section \ref{sec:one_loop}.
The interacting spins take values in space of tangent vectors to the semi-symmetric
background. We shall also evaluate the general formulas
for fields involving two derivatives. In this case the dilation operator
can be diagonalized easily so that we can read off the anomalous dimensions.

The structure of the paper is as follows. In section \ref{sec:setup} we
will recall some facts of sigma models on semi-symmetric spaces, focusing
on the one-loop conformal case that appears as part of the action in
hybrid and pure spinor models. We will also present the one-loop expansion
of the action. In section \ref{sec:fields} we briefly describe how to build
fields for the model and we spell out the leading terms of these vertex
operators in the background field expansion. Section \ref{sec:summary}
contains the main result of this work. There we describe the
full one-loop dilation operator and we analyze the anomalous dimensions
for a particular subset of fields. The results are then derived in the
section \ref{sec:one_loop}. Auxiliary integral formulas, finally, are
collected in an appendix at the end of the paper.

\section{Sigma Models on semi-symmetric spaces}\label{sec:setup}

The goal of this section is to introduce and discuss the models we
are dealing with. In the first subsection we construct their action.
Their background field expansion is the subject of the second
subsection where we present all terms that are relevant for our
one-loop computation of anomalous dimensions.

\subsection{Semi-symmetric spaces and the coset action}

The models we are about to discuss belong to the class of sigma models on coset
superspaces $G/H$ with couplings chosen such that the beta-function
vanishes, at least to leading order. Symmetric spaces, i.e. coset
spaces $G/H$ for which the Lie subalgebra $\fh\subset \fg$ is fixed
by a Lie algebra homomorphism $\sigma$ of order two, are the most
commonly considered cases, at least for bosonic groups $G$. When $G$ is
a supergroup, it is natural to consider so-called semi-symmetric
spaces in which the denominator Lie (super-)algebra $\fh\subset \fg$
is kept fixed by a homomorphism $\sigma:\fg\rightarrow \fg$ of order
four, i.e.\ $\sigma^{4}=\id$. Such a homomorphism determines a
decomposition of the form
\beq
\fg=\bigoplus_{\A=0}^{3}\fg_\A, \qquad \mbox{ where }\quad
\sigma|_{\fg_\A}=\exp\left( \frac{\pi i}{2}\A\right)\id\ .
\eeq
We shall also assume that $\fg_\A$ is contained in the
bosonic subalgebra $\fg_{\bar 0}$ for $\A=0,2$ while the subspaces $\fg_{\A}$
for $\A=1,3$ are contained in the odd part $\fg_{\bar{1}}$ of the
algebra. As we have reviewed in the
introduction, semi-symmetric superspaces of this type play an important
role, in particular  in the AdS/CFT correspondence.

Before we move on to the action of our sigma models, let us introduce
some more notation. It will be convenient to define the shorthands
$\fh\colonequals \fg_0$ and $\fm\colonequals
\bigoplus_{\A=1}^{3}\fg_\A$. Furthermore, to each component $\fg_\A$ we
can associate a projector $P_\A:\fg\rightarrow \fg_\A$. For the
projector onto the complement of $\fh$ we introduce the symbol
\beq
P=1-P_0=\sum_{\A=1}^{3}P_\A\
\eeq
in accordance with \cite{Candu:2013cga}. Let us finally fix an invariant
bilinear form $(\cdot, \cdot)$ on $\fg$. We note that any such form satisfies
\beq
\form{X}{Y}=0\quad \text{ for } \quad X\in \fg_\A, Y\in \fg_\B
\quad \mbox{ and } \quad  \A+\B\neq 0_{\text{mod}4}\ .
\eeq
This fundamental property of the invariant form will be used frequently
in what follows.

The action of our sigma models is defined on maps $g:\Sigma \rightarrow G$
from the 2-dimensional world-sheet $\Sigma$ to the supergroup $G$
\cite{Kagan:2005wt}
\beq
\label{eq:actionKaganYoung}
\calS[g]=\frac{R^{2}}{2}\int_{\Sigma}\frac{d^2z}{\pi}
\sum_{\A=1}^{3}\left(p_\A+q_\A\right)\form{P_\A\jmath}{P_{\A^{\prime}}\bar{\jmath}}\ ,
\eeq
where $\A^{\prime}\colonequals 4-\A$ and the currents $\jmath$ and $\bar{\jmath}$
are defined as
\beq \jmath = g^{-1} \partial g \quad ,\quad \bar{\jmath} = g^{-1} \bar \partial g\ .
\eeq
Note that the action $\calS$ is invariant under right multiplication $g \rightarrow g h$ by an $H$-valued function $h$ since $P_\A \fh = 0$ for $\A =
1,2,3$. By construction, the model is also invariant under a global left action of $G$, regardless of  how we choose the coefficients (couplings) $p_\A$ and $q_\A$. Reality of the action requires that
\beq
\label{eq:actioncoefficientssymmetryw}
p_\A=p_{\A^{\prime}}\quad \mbox{ and}  \qquad q_\A=-q_{\A^{\prime}}\ .
\eeq
Integrability and the vanishing of the one-loop beta-function
imposes strong constraints on the couplings. These have been worked out in \cite{Berkovits:1999im,Berkovits:1999zq,Kagan:2005wt}. In our analysis, we
shall adopt the following solution
\beqa
\label{eq:KY1loopconformal}
&p_0=q_0=0 ,&\nonumber\\[2mm]
p_\A=1\ ,& q_\A =1-\frac{\A}{2}\ , &\text{for } \A=1,2,3
\eeqa
This choice reproduces the supercoset sigma models that appear
in the pure spinor and hybrid formalisms for string  theory on
AdS spaces \cite{Berkovits:1999im}.

\subsection{One-loop action in background field expansion}

For our one-loop computation of anomalous dimensions we
need to expand the action to leading order in the background
field expansion. In order to do so, we introduce the
coordinates
\beq
\imath: G/H\rightarrow G\ ,\qquad g_0e^{\phi}\H\mapsto g_0e^{\phi}\ ,
\eeq
where $\phi\in \fm$. The expansion of the
currents $\jmath$ in these coordinates is
\beq
\label{eq:currentbackground}
\jmath=Pe^{-\phi}\partial e^{\phi} =P\left[\partial\phi-\frac{1}{2}\com{\phi}{\partial\phi}+
\frac{1}{6}\com{\phi}{\com{\phi}{\partial \phi}}\right]+\cdots\
\eeq
and similarly for $\bar\jmath$.
Let us introduce the notation $\phi_\A\colonequals P_\A\phi=t_a^{\A}
\phi^a_\A$, where $t^{\A}_a$ with $a=1,\dots,{\dim \fg_\A}$ denotes a
basis of $ \fg_\A$. Note that the objects $\phi_\A$ are Grassmann
even by construction. Hence, in working with $\phi_\A$ we do not
have to worry about the grading.

The projected currents $P_\A\jmath$ that appear in the action can now
be rewritten as
\beq
\label{eq:currentbackground2}
P_\A\jmath=\partial\phi_\A-\frac{1}{2}\sum_{\substack{\phantom{0}\\[-1mm]
\B+\C\equiv \A\\[1mm]\B,\C\neq 0}}\com{\phi_\B}{\partial\phi_\C}+\frac{1}{6}
\hspace*{-3mm}\sum_{\substack{\phantom{-}\\[-1mm]\B+\C+\D\equiv \A\\[1mm]
\B,\C,\D\neq 0}}\hspace*{-3mm}\com{\phi_\B}{\com{\phi_\C}{\partial\phi_\D}}+\cdots
\eeq
Inserting this expression into the action \eqref{eq:actionKaganYoung}
and expanding $\calS \sim \calS_0 + \calS_1$ up to the leading non-trivial
order in the coupling we obtain
\beq
\label{eq:calS0general}
\calS_0=\frac{R^{2}}{2}\int_{\Sigma}\frac{d^2z}{\pi}\sum_{\A=1}^{3}p_\A
\form{\partial \phi_\A}{\bar{\partial}\phi_{\A^{\prime}}}
\eeq
for the tree level (free) action and $\calS_1 = \calS^s_1 +\calS_1^a$ where
the symmetric part of the one-loop interaction is given by
\beqa
\label{eq:calS1sgeneral}
\calS_1^s&=&\frac{R^{2}}{2}\int_{\Sigma}\frac{d^2z}{\pi}
\Big[\hspace*{-3mm}\sum_{\substack{\phantom{-}\\[-1mm]\A+\B+\C\equiv 0}}
\hspace*{-3mm}\frac{p_\A}{2}\big\{-\form{\partial\phi_\A}
{\com{\phi_\B}{\bar{\partial}\phi_\C}}-
\form{\com{\phi_\B}{\partial\phi_\C}}{\bar{\partial}\phi_\A}\big\}\nonumber\\[2mm]
&& \hspace*{-1cm} + \hspace*{-3mm}
\sum_{\substack{\phantom{-}\\[-1mm] \A+\B+\C+\D\equiv 0}}
\hspace*{-3mm} \frac{p_\A}{6}\big\{\form{\partial\phi_\A}{  \com{\phi_\B}
{\com{\phi_\C}{\bar{\partial}\phi_\D} }}+ \form{  \com{\phi_\B}
{\com{\phi_\C}{\partial\phi_\D} }}{\bar{\partial}\phi_\A}\big\}
\nonumber\\[2mm]
&&+\sum_\A \ \sum_{\substack{\phantom{-}\\[-1mm]\B+\C\equiv \A\\[1mm]
\D+\E\equiv \A^{\prime}}}
\frac{p_\A}{4}\form{\com{\phi_\B}{\partial\phi_\C}}{\com{\phi_\D}
{\bar{\partial}\phi_\E}}\Big]\ ,
\eeqa
while the antisymmetric part takes the form
\beqa
\label{eq:calS1ageneral}
\calS_1^a&=&\frac{R^{2}}{2}\int_{\Sigma}\frac{d^2z}{\pi}\Big[\hspace*{-3mm}
\sum_{\substack{\phantom{-}\\[-1mm]\A+\B+\C\equiv 0}}
\hspace*{-3mm}\frac{q_\A}{2}\big\{-\form{\partial\phi_\A}{\com{\phi_\B}
{\bar{\partial}\phi_\C}}+\form{\com{\phi_\B}{\partial\phi_\C}}{\bar{\partial}\phi_\A}
\big\}\nonumber\\[2mm]
&&\hspace*{-1cm} + \hspace*{-3mm}
\sum_{\substack{\phantom{-}\\[-1mm] \A+\B+\C+\D\equiv 0}}
\hspace*{-3mm}\frac{q_\A}{6}\big\{\form{\partial\phi_\A}{  \com{\phi_\B}
{\com{\phi_\C}{\bar{\partial}\phi_\D} }}- \form{  \com{\phi_\B}
{\com{\phi_\C}{\partial\phi_\D} }}{\bar{\partial}\phi_\A }\big\}\nonumber\\[2mm]
&&+\sum_\A \sum_{\substack{\phantom{-}\\[-1mm]\B+\C\equiv \A \\\D+\E\equiv \A^{\prime}}}\frac{q_\A}{4}
\form{\com{\phi_\B}{\partial\phi_\C}}{\com{\phi_\D}{\bar{\partial}\phi_\E}}\Big]\ .
\eeqa
From the tree level action $\calS_0$ in eq.\ \eqref{eq:calS0general} we read off
that the free two-point correlation function is given by
\beq\label{eq:freecorrelator}
\vac{\phi_\A(z,\bar{z})\otimes\phi_{\A^{\prime}}(w,\bar{w})}_0=
-\frac{(-1)^{|a|}}{R^{2}p_\A}\log\left|\frac{z-w}{\epsilon}\right|^2
\sum_{a=1}^{\dim\fg_\A}t_{a}^{\A}\otimes t^{\A^{\prime}}_{a'}\eta^{aa'}\ .
\eeq
In our conventions $(t^b,t^a)=\eta^{ab}$ and $t^a=\eta^{ab}t_b$.
In passing we note that the two point function between a holomorphic and
anti-holomorphic derivative of the fundamental fields contains a contact
term,
\beq\label{eq:contact}
\vac{\partial\phi_\A(z,\bar{z})\otimes \bar\partial\phi_{\A^{\prime}}
(w,\bar{w})}_0= \frac{(-1)^{|a|}}{R^{2}p_\A} \delta^{2}(z-w)
\sum_{a=1}^{\dim\fg_\A}t_{a}^{\A}\otimes t^{\A^{\prime}}_{a'}\eta^{aa'}\ .
\eeq
These need to be taken into account if one or both of the insertion
points are integrated over, see below.

In our analysis we shall split the one-loop terms of the interaction
into three vertices and four vertices,
\begin{align}
\calS_{1}=\int \frac{d^2z}{\pi}\left(\Omega_3(z,\bar{z})+
\Omega_4(z,\bar{z})\right)\ .
\end{align}
Once again, we can then split these vertices into a symmetric and an
antisymmetric part, i.e.\ $\Omega_p = \Omega_p^s +  \Omega_p^a$. If we
consider the one-loop conformal case \eqref{eq:KY1loopconformal}, the
previous expressions simplify drastically. In particular the first row
of eq.\ \eqref{eq:calS1sgeneral} cancels so that
\beq
\Omega^s_3(z,\bar{z})=0
\eeq
and the first row of eq.\ \eqref{eq:calS1ageneral} is
\beqa
\label{eq:Omega3}
\Omega_3^a=-\sum_{\alpha+\beta+\gamma=0}\frac{R^2}{4}(q_\alpha+q_\gamma)
(\partial\phi_\alpha,[\phi_\beta,\bar{\partial}\phi_\gamma])\ .
\eeqa
For some of the calculation we go through later is will be useful to
see this interaction explicitly,
\beqa
\label{eq:omega3a}
\Omega^a_3(z,\bar{z})&=&\frac{R^2}{2}\big\{-\frac12(\partial \phi_1,[\phi_2,\bar\partial\phi_1])
+\frac12(\partial \phi_3,[\phi_2,\bar\partial\phi_3]) \nonumber \\[2mm]
& & -\frac14(\partial \phi_2,[\phi_1,\bar\partial\phi_1])
+\frac14(\partial \phi_2,[\phi_3,\bar\partial\phi_3]) \\[2mm]
& & -\frac14(\partial \phi_1,[\phi_1,\bar\partial\phi_2])+
\frac14(\partial \phi_3,[\phi_3,\bar\partial\phi_2])\big\}\ . \nonumber
\eeqa
In some calculations we will modify the vertex $\Omega_3$ by adding a
total derivative. This can simplify the expression significantly. One
example is given by the following modification of the three vertex,
\beqa
\Omega_3^{\prime a} \equiv  \Omega_3(z,\bar{z})^a+
\frac{R^2}{8}\partial\big( (\phi_2,[\phi_1,\bar\partial\phi_1])
- (\phi_2,[\phi_3,\bar\partial\phi_3])\big)\nonumber\\[2mm]
+\frac{R^2}{8}\bar\partial\big( (\partial\phi_1,[\phi_1,\phi_2])-
(\partial\phi_3,[\phi_3,\phi_2])\big)
\nonumber\\[2mm]=-\frac{R^2}{2}(\partial\phi_1,[\phi_2,\bar \partial\phi_1])
+\frac{R^2}{2}(\partial\phi_3,[\phi_2,\bar \partial\phi_3])\ .
\eeqa
In order to spell out the four vertex $\Omega_4$ we make use of the
following simple equality
\beqa
(\partial \phi_\A,[\phi_\B,[\phi_\C,\bar\partial\phi_\D]])=([\phi_\C,[\phi_\B,\partial\phi_\A]],\bar\partial\phi_\D)
=-([\phi_\B,\partial\phi_\A],[\phi_\C,\bar\partial\phi_\D])\,,
\eeqa
This allows to bring $\Omega_4$ into the form
\beqa
\label{eq:Omega4}
\Omega_4=\frac{R^2}{2}\!\!\!\sum_{\A+\dots+\D=0}\!\!
\left(\frac{p_{\A+\B}}4-\frac{p_\A}{3} -\frac {q_\A-q_\D}{6}+\frac{q_{\A+\B}}4\right)\form{[\phi_\B,\partial\phi_\A]}
{[\phi_\C,\bar\partial\phi_\D]}\,
\eeqa
which concludes our discussion of the action and in particular the
various interaction terms.

\section{Coset fields and their expansion}\label{sec:fields}

Our discussion of fields splits into two parts again. First we will
review the construction of vertex operators for coset spaces $G/H$,
see \cite{Candu:2013cga,Cagnazzo:2014yha}. Then, we expand these
operators up to the first non-trivial order in the background field
expansion.

\subsection{Vertex operators in coset models}

As explained in \cite{Candu:2013cga,Cagnazzo:2014yha}, vertex operators
for coset models are composed of a zero mode contribution and a tail
of currents. The latter contains all the derivatives.

The zero mode factor of a vertex operator is associated with a square integrable
section of a homogeneous vector bundle over the coset. The bundle depends
on the tail of currents, see below. Following the notation used in \cite{Candu:2013cga},
we will denote the space of sections by $\Gamma_\lambda$, where $\lambda$ labels a representation of the denominator subgroup $H \subset G$ and hence a homogeneous
vector bundle over $G/H$. The space of such sections can be obtained as
\begin{equation}
\Gamma_\lambda = \Gamma_\lambda(G/H) = \{ F \in L_2(G) \otimes {\cal S}_\lambda  :
F(gh) = S_\lambda(h^{-1}) F(g)\
\forall h\in H\}\
\end{equation}
where ${\cal S}_\lambda$ denotes the carrier space of the representation $S_\lambda$
of  $H$. For the associated vertex operators we use the symbol
\begin{eqnarray}
\label{eq:vertexoperators}
{V}_{\Lambda\lambda}(z,\bar{z})\defeq V[{\cal D}_{\Lambda\lambda}](z,\bar z)
\quad \mbox{ for } \quad  {\cal D}_{\Lambda\lambda} \in \Gamma_\lambda
\ .
\end{eqnarray}
Let us note that the space of sections $\Gamma_{\lambda}$ of a
homogenous vector bundle on $G/H$ carries an action of the global
symmetry (super-)group $G$.
Under this action, the space $\Gamma_\lambda$ decomposes
into irreducible representations. The latter are labeled by $\Lambda$.
How many times any given $\Lambda$ appears within $\Gamma_\lambda$ can only be
determined with the tools of harmonic analysis. When both $G$ and $H$ are
compact, there exists a universal formula for the multiplicities, see
\cite{Candu:2013cga}. However, the counting of fields is an issue
that can be treated separately from the computation of their anomalous
dimensions.

In addition to the zero mode factor, vertex operators also contain a
tail factor composed
of currents and their derivatives. For these we introduce the compact notation
\beq\label{eq:composite_field}
\jmath_{\tb m}(z,\bar{z})=
\bigotimes_{\rho=1}^r\partial^{m_\rho-1} \jmath(z,\bar{z})\ ,
\eeq
with a multi-index  $\tb{m}=\{m_1,\ldots, m_r\}$ such that $m_\rho \geq
m_{\rho+1}\geq 1$. The definition of $\bar\jmath_{\bar{\tb m}}$ is
analogous. Note that these composite fields transform in (a
subrepresentation of)
the tensor product of the $r$-fold tensor product of the $\fh$
representation $\fm = \fg/\fh$.

Let us now assemble the complete vertex operators from the zero mode
factor and the tail of currents as follows
\beq
\label{eq:fieldscosetmodel}
 \Phi_{\boldsymbol{\Lambda}}(z,\bar z) = {\sf d}_{\lambda\mu\bar\mu}
(V_{\Lambda\lambda}\otimes
\jmath_{\tb{m}}\otimes \bar \jmath_{\widebar{\tb{m}}})(z,\bar z) \ , \qquad \boldsymbol{\Lambda}\defeq (\Lambda, \lambda,\mu,\bar{\mu})\ .
\eeq
This formula contains one new ingredient, namely the intertwiner ${\sf d}$
which ensures that $\Phi_{\boldsymbol{\Lambda}}$ is actually invariant under
global $H$ transformations, as it has to in order for it to represent a
physical operator of the $G/H$ coset model. The intertwiner ${\sf d}$ is
itself composed of several pieces,
\beq
{\sf d}_{\lambda\mu\bar\mu}\defeq {\sf c}_{\lambda\mu\bar\mu}(1\otimes \mathsf{p}_{\mu}\otimes \mathsf{p}_{\bar \mu})\ ,
\eeq
Here, ${\sf c}$ denotes an intertwiner between the triple tensor product of
the representations $[\lambda],\, [\mu]$ and $[\bar \mu]$ of the denominator
algebra $\fh$ and the trivial representation on the complex numbers, i.e.\
\begin{equation}\label{C}
{\sf c}^{\lambda\mu\bar \mu}: [\lambda]\otimes [\mu]
\otimes [\bar \mu]  \ \rightarrow\ \mathbb{C}\ .
\end{equation}
The maps ${\mathsf p}$ project from a (partially symmetrized)
tensor power of the representation $\fm = \fg/\fh$ down to the irreducible
representations $\mu$ and $\bar\mu$ of the denominator algebra $\fh$.

In our discussion below we shall often suppress the $\fg$ and $\fh$
representation labels $\Lambda, \lambda, \mu, \bar\mu$ and simply write
\beq
\Phi_{\boldsymbol{\Lambda}}(z,\bar{z})={\sf d}\, \circ \,V\otimes\jmath_{\tb m}\otimes\jmath_{\widebar{\tb{m}}}\ .
\eeq

\subsection{Background field expansion of vertex operators}

The one-loop expansion of a general coset field around an arbitrary
point $g_0H$ may be written schematically as
\beq
\label{eq:1loopfields}
\Phi_{\Lambda}(z,\bar{z}\mid g_0)={\sf d}\, \circ\, \big(V^{(0)}+V^{(1)}
\ldots\big)\otimes \big(\jmath_{\tb m}^{(0)}+\jmath_{\tb m}^{(1)}\ldots\big)
\otimes\big( \bar \jmath_{\widebar{\tb{m}}}^{(0)}+\bar \jmath_{\widebar{\tb{m}}}^{(1)}\cdots\big)\ ,
\eeq
Let us spell out concrete expressions for the various terms in the expansion.
For the zero mode contribution $V_{\Lambda\lambda} = V$ the leading terms in
the background field expansion read
\begin{eqnarray}
V& = & \calD_{\Lambda\lambda}(g_0)-L_\Lambda(\mathrm{Ad}_{g_0}\phi(z,\bar{z}))
\calD_{\Lambda\lambda}(g_0)+\dots\\[2mm]
& = & V^{(0)}+\sum_{\alpha=1,2,3}V^{(1)}(t_a)\phi_\alpha^a+\dots\ .
\end{eqnarray}
Here, $L_\Lambda(X)$ denotes the right invariant vector field that is
associated with an element $X \in \fg$, see \cite{Candu:2013cga} for more
details. By definition, $V^{(0)}$ is the constant term in the expansion
around $g_0$.

Similarly, we can also expand the tail of currents. For the current $\jmath$
one finds that
$$ \jmath = \partial \phi -  \frac{1}{2}
\sum_{\substack{\phantom{-}\\[-2mm]
\B+\C\neq 0\\[1mm]\B,\C\neq 0}}\com{\phi_\B}{\partial\phi_\C} + \dots
\ . $$
Note that in the case of symmetric spaces, i.e.\ when the index $\A$ runs
over $\A=0,1$ only, the leading nontrivial term vanishes because the sum
of $\B=1$ and $\C=1$ is $\B+\C=0$ mod $2$. Hence, while this term did not
appear in \cite{Candu:2013cga}, we need to consider it in dealing with
semi-symmetric spaces. When the expansion of the current is inserted
into our expression \eqref{eq:composite_field} it gives
\beq
\begin{split}
\label{eq:defproductcurrents2}
\jmath_{\tb m}&\colonequals \jmath_{\tb m}^{(0)}+\jmath_{\tb m}^{(1)}+\dots
= \ \bigotimes_{\rho=1}^r \partial^{m_\rho}\phi + \nonumber \\[2mm] &  -\sum_{k=1}^r\bigotimes_{\rho=1}^{k-1} \partial^{m_\rho}\phi\otimes\partial^{m_{k}-1}\left[\frac{1}{2}
\sum_{\substack{\phantom{-}\\[-2mm]\B+\C\neq 0\\[1mm]\B,\C\neq 0}}\com{\phi_\B}{\partial\phi_\C}\right]
\otimes \bigotimes_{\rho=k+1}^{r} \partial^{m_\rho}\phi\dots \nonumber\,
\end{split}
\eeq
and similarly for the tail factors $\bar\jmath_{\widebar{\tb{m}}}$ in which
all the unbarred labels $m$ are replaced by barred ones and antiholomorphic
derivatives $\bar \partial$ appear instead of the holomorphic ones.

\section{One-loop dilation operator: Summary of results}
\label{sec:summary}

Since the computation of the one-loop dilation operator is a bit involved, we
shall state our results beforehand and also apply the formula to one set of
operators that includes the sigma model interaction. This will allow us to
re-derive the vanishing of the beta-function for the models under
consideration.

\subsection{The one-loop dilation operator}

It is useful to begin with a short description of the space of states
which our dilation operator acts upon. As one may infer from the discussion
in the previous section, this space is constructed out of two building blocks.
To begin with, the zero mode term in our vertex operators contributes a factor $L_2(G)$ of square integrable functions on $G$. Recall that the space $L_2(G)$ carries two commuting actions of $\mathfrak{g}$. After restricting the right action from $\mathfrak{g}$ to the subalgebra $\mathfrak{h}$ we obtain
$$ L_2(G) \cong \bigoplus_\lambda \Gamma_\lambda(G/H) \otimes
\mathcal{S}_\lambda \  $$
which describes the decomposition of $L_2(G)$ into $\mathfrak{g}$ left
and $\mathfrak{h}$ right modules. The second building block of the state
space is the familiar $\mathfrak{h}$ module $\mathfrak{m}  = \mathfrak{g}/\mathfrak{h}$ which comes with the currents. From these
two elements we can construct
\begin{equation} \label{Hrr}
{\mathcal H}_{r,\bar r} \equiv \left(L_2(G) \otimes \mathfrak{m}^{\otimes_r}
  \otimes \mathfrak{m}^{\otimes_{\bar r}}\right)^\mathfrak{h}
= \bigoplus_{\lambda} \Gamma_\lambda(G/H)\otimes
\left(\mathcal{S}_\lambda \otimes\mathfrak{m}^{\otimes_r}
  \otimes \mathfrak{m}^{\otimes_{\bar r}} \right)^{\mathfrak{h}}
\end{equation}
where the superscript $\mathfrak{h}$ means that we project to the space of
$\mathfrak{h}$ invariants. The space of all vertex operators with naive scaling
dimension $(h_0,\bar h_0) = (n,\bar n)$ has the form
\begin{equation}\label{sof}
{\mathcal H}^{(n,\bar n)} = \bigoplus_{r,\bar r}
\left[ \sum_{l=0}^{n-r} p_l(n-r)+ \sum_{l=0}^{\bar n -\bar r}
p_{l}(\bar n-\bar r) \right] {\mathcal H}_{r,\bar r}
\end{equation}
where $p_l(m)$ denotes the number of partitions of $m$ that have length
$l$. These multiplicities arise from the placement of derivatives in the
tail of currents.

We will now construct the 1-loop dilation operator from an $\mathfrak{h}$ invariant operator on the space $L_2(G) \otimes \mathfrak m^r \otimes
\mathfrak{m}^{\bar r}$. It consists of three separate pieces,
\begin{equation}\label{dilation}
D^{\text{1-loop}}_{r,\bar r} = H^\text{h} + H^\text{ht} + H^\text{tt}\ .
\end{equation}
The first term acts on the zero mode factor only and it is identical to the
so-called Bochner Laplacian for sections of vector bundles,
$$ H^\text{h} = \Cas_\mathfrak{g} - \Cas_\mathfrak{h}\ . $$
Let us stress that this operator acts only on $L_2(G) \cong \bigoplus_\lambda
\Gamma_\lambda(G/H) \otimes \mathcal{S}_\lambda$ where $\cong$ is an isomorphism
of a $\mathfrak{g}$ left and an $\mathfrak{h}$ right module. In particular, the second term acts through the right action of $\mathfrak{h}$ on functions.

The second term in our dilation operator \eqref{dilation} describes an
interaction between the tail of currents and the head. It reads
$$ H^\text{ht} = - \sum_{\rho=1}^{r}   ((-1)^c+Q_{\A\gamma}) R(t_c) \left(t^c\right)^\rho_\A
- \sum_{\bar \rho=1}^{\bar  r} ((-1)^c+Q_{\A\gamma}) R(t_c) \left(t^c\right)^{\bar \rho}_\A $$
where $R(t_c)$ denotes the right action of elements $t_c \in \mathfrak{g}$ on
$L_2(G)$. The indices on $t_c \in \mathfrak{g}_\gamma$ indicate that the generator
acts on a component of the $\rho^\text{th}$ current that lies in the subspace
$\mathfrak{g}_\A \subset \mathfrak{m}$. Finally, we have also introduced
\begin{equation}\label{eq:Q}
Q_{\alpha\beta} \equiv q_\alpha+q_\beta-q_{\alpha+\beta} \ .
\end{equation}
Note that the action of $H^\text{ht}$ on vertex operators induces transitions
between different vector bundles on $G/H$.

The most interesting contribution is the tail-tail interaction  term. It consists
of a sum of pairwise interactions between left- and right moving currents. The
interaction terms resemble the XXZ Hamiltonian for spin-spin interaction,
at least when written in an appropriate form. Let us recall that the XXZ
Hamiltonian can be written as
$$ H^\text{XXZ}_\text{pair} = \sigma^x \otimes \sigma^x + \sigma^y \otimes \sigma^y +\Delta\sigma^z \otimes \sigma^z = (1+\Delta) (I+P) + (1-\Delta) K
\ . $$
Here, the $\sigma^a$ are Pauli-matrices, $I$ is the identity on the tensor
product of two spin representations, $P$ the permutation and $K$  the
projection to the spin zero subspace.
Similarly, we now define three operators on the tensor product $\mathfrak{m}
\otimes \mathfrak{m}$ that are somewhat similar to $I,P$ and $K$,
\begin{eqnarray}
X^{abcd}_{\alpha\beta\gamma\delta} & = & (-1)^{|c||d|} f^{bai}{f^{cd}_{\phantom{cd}i}}
\ , \label{eq:X}\\[2mm]
A^{abcd}_{\alpha\beta\gamma\delta} & = & (-1)^{|c|} f^{adi}{f^{cb}_{\phantom{cd}i}}\ ,\label{eq:A}\\[2mm]
H^{abcd}_{\alpha\beta\gamma\delta} & = & (-1)^{|b|+|b||d|} f^{aci}{f^{bd}_{\phantom{cd}i}}\ .\label{eq:H}
\end{eqnarray}
Here, the Greek letters run through $\alpha,\dots=1,2,3$ and the corresponding
Roman letters $a, \dots$ run through a basis of $\mathfrak{g}_\alpha,\dots$.
\footnote{To be fully precise we should actually let Greek indices run over
irreducible subrepresentations of the $\mathfrak{h}$ action on $\mathfrak{m}$
and the corresponding Latin indices over a basis of these irreducible subspaces.
We think of these additional (finer) projections as being implemented implicitly.}
Let us also stress that the index $i$ runs over a basis of $\mathfrak{g}$.
In terms of the operators $X,A$ and $H$, we can write tail-tail interaction
terms as
\begin{eqnarray*}
 H^\text{tt} & = & \sum_{\rho,\bar\rho = 1}^{r,\bar r} \sum_{\alpha,\dots,
 \delta = 1}^3 \frac12 \left(p_{\alpha+\beta}
   + (-1)^{|a|+|c|} Q_{\alpha\beta}Q_{\gamma\delta}\right)
   X^{\rho\bar \rho}_{\alpha\beta\gamma\delta} +
   \frac12 \left(p_{\alpha+\delta} +\phantom{\sum}  \right. \\[2mm]
   & & \hspace*{-.8cm}
   \left. + (-1)^{|a|+|c|} Q_{\alpha\delta}Q_{\gamma\beta}\right)
   A^{\rho\bar \rho}_{\alpha\beta\gamma\delta}
   - \frac12 \left(p_{\alpha+\gamma}
   + (-1)^{|a|+|c|} Q_{\alpha\gamma}Q_{\beta\delta}\right)
   H^{\rho\bar \rho}_{\alpha\beta\gamma\delta}\\[4mm]
   & & \hspace*{1cm}
   +\left(2+ Q_{\alpha\gamma}
    -  Q_{\beta\delta}\right)
   H^{\rho\bar \rho}_{\alpha\beta\gamma\delta}
\end{eqnarray*}
Note that all three contributions to the operator \eqref{dilation} commute
with the action of $\mathfrak{h}$ on $L_2(G) \otimes \mathfrak{m}^{r+\bar r}$
and hence the sum descends to a well-defined operator on the space \eqref{sof}
of vertex operators in the sigma model.

Let us conclude with two remarks on the validity of our formula \eqref{dilation}
of the dilation operator. We want to stress that the expressions we provided also
hold for non-compact semi-symmetric spaces, such as AdS geometries. In those cases,
the zero mode spectrum is continuous and that allows us to analytically continue the
results beyond $L_2(G)$ to some non-normalizable functions on the
target space. Thereby, the formulas can also be applied, e.g.\ to the Noether
currents of the global $\mathfrak{g}$ symmetry which involve functions on the
target space that transform in the adjoint representation.

Finally, while the formulas are derived for semi-symmetric spaces, they can
also be applied to symmetric spaces in which the denominator algebra $\mathfrak{h}$
is left invariant under an automorphism of order two rather
than four. In that case, only the zero model Hamiltonian $H^\text{h}$ and
one term from the tail-tail interaction $H^\text{tt}$, namely the term
$2 H^{\rho\bar \rho}_{\alpha\beta\gamma\delta}$, can contribute, and we
recover what has been found in \cite{Candu:2013cga}. Let us stress that
our analysis here assumes that all elements in a subspace $\mathfrak{g}_
\alpha$ are either bosonic or fermionic. The investigations in
\cite{Candu:2013cga} were a bit more general in that they did not impose
such a condition. In fact, for many interesting symmetric spaces the
subspace $\mathfrak{g}_1$ contains both fermionic and bosonic
directions.

\subsection{Example: Marginal operators}

As an example, let us consider the subspace of operators of naive dimension
$(h,\bar h) = (1,1)$ whose head is a constant function on the semi-symmetric
target space. In this case, the contributions from $H^\text{h}$ and $H^{\text{ht}}$
vanish trivially. This implies that the space of such operators does not mix with
operators whose zero mode factors are non-trivial. To be even more concrete let
us pick the example of $AdS_5\times S^5$ in which the space $\mathfrak{m}$
decomposes into four irreducible representations of the denominator subgroup.
These are the fermionic subspaces $\mathfrak{g}_1$ and $\mathfrak{g}_3$ along
with the direct summands of $\mathfrak{g}_2=\mathfrak{g}_{2^+} \oplus
\mathfrak{g}_{2^-}$ that are associated with tangent vectors of the sphere
$S^5$  and of $AdS_5$, respectively. It follows that the space of $\mathfrak{h}$
invariant operators is spanned by the following four basis elements
\begin{equation} \label{eq:basis}
(\form{\jmath_1}{\bar\jmath_3},\form{\jmath_{2^+}}{\bar\jmath_{2^+}},
\form{\jmath_{2^-}}{\bar\jmath_{2^-}},\form{\jmath_3}{\bar\jmath_1})\
\end{equation}
From these fields of naive scaling dimension $(h_0,\bar h_0)=(1,1)$ one can
build the kinetic operator
\beqa
\label{eq:kin}
\mathcal{L} \equiv \frac32 \form{\jmath_1}{\bar\jmath_3}+
\form{\jmath_{2^+}}{\bar\jmath_{2^+}}+
\form{\jmath_{2^-}}{\bar\jmath_{2^-}}+\frac 12 \form{\jmath_3}{\bar\jmath_1}\ .
\eeqa
This is the combination of marginal operators that appears in the integrand
of the action. In order for the $AdS_5\times S^5$ sigma model to be one-loop
conformal, the operator $\mathcal{L}$ must possess vanishing one-loop anomalous
dimension. Using the results we sketched in the previous subsection we shall
verify that $\mathcal{L}$ is an eigenvector of our one-loop dilation operator
with zero eigenvalue.

In the basis \eqref{eq:basis}, our one loop dilation operator acts as a
$4 \times 4$ matrix whose elements are quadratic in the structure constants.
The matrix elements can easily be worked out from the general formula for
$H^\text{tt}$ with $\rho =1 = \bar \rho$. Each matrix element corresponds
to a specific configuration of the parameters $\A,\B,\C,\D$. After inserting
the values of $p_\A$ and $q_\A$ from eq.\ \eqref{eq:KY1loopconformal} it
turns out that the matrix elements are all given by $\pm C$ where $C$ is the
value of the Casimir element $\Cas_\mathfrak{h}$ in the representation
$\mathfrak{g}_{2^+}$ of the denominator algebra $\mathfrak{h}$. More precisely,
using the relation
\beqa
f_{ad}^{\phantom{ad}e}f_{bc}^{\phantom{cb}d}\eta^{ab}=0\,
\eeqa
which holds for all Lie superalgebras $\mathfrak{g}$ with vanishing dual
Coxeter number,  one finds that
\begin{displaymath}
D^\text{1-loop}_{1,1} \ = \ C \,\cdot \,
\left( \begin{array}{cccc}
0 & 1 & -1&0 \\
-1 & 1 & 0&1 \\
1 & 0 & -1 &-1\\
0&-1&1&0
\end{array} \right)\ .
\end{displaymath}
It is now easy to verify that the vector $(3/2,1,1,1/2)$ is indeed an
eigenvector of the dilation operator with vanishing eigenvalue, as we
anticipated.

Similarly one can also analyze the anomalous dimensions of operators
$V_{\Lambda\mathfrak{m}} \jmath$ with naive scaling weight $(h_0,\bar h_0)
= (1,0)$ with $\Lambda$ the adjoint representation of $\mathfrak{g}$.
Once again there are four vertex operators of this type and the
dilation operator can be worked out easily. In this case it receives
contributions from the Bochner Laplacian and the head-tail interaction
$H^\text{ht}$. The dilation operator can be shown to possess an
eigenvector with zero eigenvalue, corresponding to a component of
the conserved Noether current of the theory.

\section{One-loop dilation operator: The derivation}
\label{sec:one_loop}

Having stated the main results we now want to go through our
calculations. These are  a lot more involved than in the case of
symmetric superspaces since the action we consider here contains
many new terms that have no analogue for NLSM on symmetric spaces.
The section begins with a brief outline of the computation. Then
we dive into the details and evaluate all the various contributing
terms.

\subsection{Outline of the computation}

The one-loop anomalous dimension $\delta \mathbf{h}= \delta\mathbf{h}_\Phi$ of a
field $\Phi$ appears in the coefficient of the logarithmic singularity of the two
point function at one-loop, see e.g.\ \cite{Candu:2013cga} for a detailed discussion,
$$
\vac{\Phi_{\boldsymbol{\Lambda}}(u,\bar{u})\otimes
\Phi_{\boldsymbol{\Xi}}(v,\bar{v})}_1 = \vac{2\delta\mathbf{h}\cdot
\Phi_{\boldsymbol{\Lambda}}(u,\bar{u})\otimes \Phi_{\boldsymbol{\Xi}}
(v,\bar{v})}_0\, \log\left|\frac{\epsilon}{u-v}\right|^2 + \cdots \ .
$$
The correlation function on the right hand side is evaluated in the free theory,
i.e.\ by performing Wick contractions with the propagator
\eqref{eq:freecorrelator}. By definition, the one-loop correlation function
on the left hand side is obtained as the leading non-trivial term in
\begin{multline}
\label{eq:defFbulk}
\vac{\Phi_{\boldsymbol{\Lambda}}(u,\bar{u})\otimes
\Phi_{\boldsymbol{\Xi}}(v,\bar{v})}=\\[2mm]
=\int_{G/H}d\mu(g_0H)\vac{\Phi_{\boldsymbol{\Lambda}}(u,\bar{u}\mid g_0)\otimes
\Phi_{\boldsymbol{\Xi}}(v,\bar{v}\mid g_0)e^{-\calS_{\text{int}}}}_{0,c}\ ,
\end{multline}
where the subscript $c$ stands for `connected'. When counting loops,
recall that each propagator carries
a factor $1/R^2$ and each insertion of the interaction produces a factor
$R^2$. The one-loop contribution contains all terms that are suppressed
by a factor $1/R^2$ relative to the tree level.

Expanding the two point correlation function \eqref{eq:defFbulk} to
one-loop, we have
\begin{equation}
\label{eq:1loopexpansion2pt}
\vac{\Phi_{\boldsymbol{\Lambda}}(u,\bar{u})\otimes \Phi_{\boldsymbol{\Xi}}(v,\bar{v})}=\int_{G/H}d\mu(g_0H)\,
{\sf d}\otimes {\sf d}\big(I_0+I_1 + \dots \big)\ ,
\end{equation}
where
\beqa
I_0=\vac{V^{(0)}\otimes \jmath_{\tb m}^{(0)}(u)\otimes \bar
\jmath_{\widebar{\tb{m}}}^{(0)}(\bar u)\otimes V^{(0)}\otimes
\jmath_{\tb m}^{(0)}(v)\otimes\bar \jmath_{\widebar{\tb{m}}}^{(0)}(\bar v)}_0\ .
\eeqa
Here we used the schematic representation \eqref{eq:1loopfields} and
the fact that derivatives of the fundamental field are
(anti)holomorphic in the free theory. In order to have a non-vanishing
tree level result, the total $\mathbb{Z}_4$ grading of all
currents $\jmath$ and their derivatives has to vanish. The same
condition must be satisfied for $\bar\jmath$ and all their
derivatives. One may note that this simple selection rule
ceases to hold at loop level, at least on the face of it.

Let us now turn to $I_1$. As we noted above, the $R^{-2}$
corrections to the correlation functions are collected in $I_1$.
There are a variety of different terms. To begin with, there are
three different terms in which no interaction term appears. In
order to produce the desired factor $1/R^2$, these must involve
one additional Wick contraction compared to the tree-level term.
This contraction can either involve the two zero mode factors (case A),
or two fields from the tails (case C) or one field from the head
and one from the tail (case G). Next, there exist several terms
that involve one interaction term. If the latter is given by the
three vertex $\Omega_3$, then we must expand either a zero mode
factor $V$ (case F) or a tail $\jmath$ (case D) to the leading
non-trivial order. Terms involving a single interaction term
$\Omega_4$ contain two additional Wick contractions compared to
the tree-level computation and hence also contain one factor
$1/R^2$ (case B). Finally, we also have to consider  one type of
contribution with two interaction three-vertices $\Omega_3$ since
these contain three additional Wick contractions compared to
tree-level (case E).

The quantity $I_1$ is obtained by summing all these different
contributions, i.e.\ $I_1 = \sum I^\nu_K$ where
\begin{eqnarray}
I_A & =& \vac{V^{(1)}_u\otimes \jmath_{\tb m,u}^{(0)}\otimes
\bar \jmath_{\widebar{\tb{m}},u}^{(0)}\otimes V^{(1)}_v\otimes
\jmath_{\tb m,v}^{(0)}\otimes\bar \jmath_{\widebar{\tb{m}},v}^{(0)}}\\[2mm]
I_B & =& \vac{V^{(0)}_u\otimes \jmath_{\tb m,u}^{(0)}\otimes \bar \jmath_{\widebar{\tb{m}},u}^{(0)}\otimes V^{(0)}_v\otimes \jmath_{\tb m,v}^{(0)}\otimes \bar \jmath_{\widebar{\tb{m}},v}^{(0)}\ {\cal O}_4}\\[2mm]
I^1_C & =& \vac{V^{(0)}_u\otimes \jmath_{\tb m,u}^{(1)}\otimes \bar \jmath_{\widebar{\tb{m}},u}^{(0)}\otimes V^{(0)}_v\otimes \jmath_{\tb m,v}^{(1)}\otimes\bar  \jmath_{\widebar{\tb{m}},v}^{(0)}}\\[2mm]
I^1_D & =&  \vac{V^{(0)}_u\otimes \jmath_{\tb m,u}^{(1)}\otimes \bar \jmath_{\widebar{\tb{m}},u}^{(0)}\otimes V^{(0)}_v\otimes \jmath_{\tb m,v}^{(0)}\otimes\bar \jmath_{\widebar{\tb{m}},v}^{(0)}\ {\cal O}_3}\\[2mm]
I_E & =& \vac{V^{(0)}_u\otimes \jmath_{\tb m,u}^{(0)}\otimes\bar  \jmath_{\widebar{\tb{m}},u}^{(0)}\otimes V^{(0)}_v\otimes \jmath_{\tb m,v}^{(0)}\otimes \bar \jmath_{\widebar{\tb{m}},v}^{(0)}\ \frac12{\cal O}^2_3}\\[2mm]
I^1_F & =& \vac{V^{(1)}_u\otimes \jmath_{\tb m,u}^{(0)}\otimes \bar\jmath_{\widebar{\tb{m}},u}^{(0)}\otimes V^{(0)}_v\otimes \jmath_{\tb m,v}^{(0)}\otimes \bar \jmath_{\widebar{\tb{m}},v}^{(0)}\ {\cal O}_3}\\[2mm]
I^1_G & =& \vac{V^{(1)}_u\otimes \jmath_{\tb m,u}^{(0)}\otimes\bar \jmath_{\widebar{\tb{m}},u}^{(0)}\otimes V^{(0)}_v\otimes \jmath_{\tb m,v}^{(1)}\otimes\bar  \jmath_{\widebar{\tb{m}},v}^{(0)}}
\end{eqnarray}
Here, the subscripts $u$ and $v$ label fields that are inserted at $(u,\bar u)$ and
$(v,\bar v)$, respectively. We have introduced the following shorthand notation for integrated interaction vertices
$$ {\cal  O}_3 = -\int_\mathbb{C} \frac{d^2z}{\pi}\  \Omega_3(z,\bar z) \quad ,\quad
   {\cal  O}_4 = -\int_\mathbb{C} \frac{d^2z}{\pi}\  \Omega_4(z,\bar z)\ .  $$
Note that we included the minus sign from the $\exp(-S)$ into our definition of
${\cal O}_{3,4}$. Similarly, we also put a factor $1/2$ into our definition of
$I_E$ because this term arises from the second order term in the expansion of
$\exp(-S)$. We have to consider $I_C^\nu$, $I_D^\nu$ and $I_G^\nu$ with $\nu
=1,2,3,4$ that are given by all the possible combination obtained by expanding
one of the elements in the first field ($V_1, \jmath_1, \bar \jmath_1$)
and one of the elements in the second field
($V_2,\jmath_2 ,\bar \jmath_2$).

We would like to stress again that, unlike in the case of symmetric
superspaces, the one loop dilation operator for semi-symmetric spaces
does no longer act diagonally on the basis of fields we have selected.
The diagonal terms of the dilation operator are obtained considering
those pairs of fields from our basis that possess a non-vanishing tree
level two-point function. We shall see that these diagonal terms are
formally identical to $\mathbb{Z}_2$ case, at least after all cancelations
are  taken into account. The existence of off-diagonal entries in the
dilation operator is quite crucial, however. As we have seen at the
end of the previous section, for example, these terms conspire to make
the one-loop anomalous dimension of the action vanish.

\subsection{Detailed calculation of cases A-G}

\subsubsection{Contributions from case A}

In this case the result is the same as for symmetric spaces. It is determined
by the logarithmic contribution from $I_A$. Since the only logarithmic term
arises from a contraction of the two dimension zero fields $\phi$, without
any derivative, the only expression that needs to be evaluated is
\beqa
\label{eq:contributionfromfieldexpansion}
&&\int_{G/H}d\mu(g_0H)\,
\vac{V^{(1)}(u,\bar u)\otimes V^{(1)}(v,\bar v)}=\\[2mm] &&=
R^{-2}\log\left|\frac{u-v}{\epsilon}\right|^2\,
\int_{G/H}d\mu(g_0H)\,
\left[\Cas_{\g}^\Lambda - \Cas_{\h}^\lambda\right]V^{(0)}\otimes V^{(0)}
\ .\nonumber
\eeqa
The details of the calculations can be found in subsection 3.2.1 of
\cite{Candu:2013cga}. Note that it is important here that we chose the
parameters $p_\A=1$. Beyond this conformal case, the results takes a more
complicated form which cannot be written in terms of  $\Cas_{\g}^\Lambda
- \Cas_{\h}^\lambda$. Hence, while it was very easy to compute the
corrections to the scaling behavior in symmetric spaces beyond the
conformal case, see \cite{Candu:2013cga}, a simple extension to
non-conformal semi-symmetric spaces does not exist.

\subsubsection{Contributions from case B}
Let us now turn to the more  interesting case B and analyze the following
integral that is contained in it
\beqa \label{eq:CorrIB}
\tilde I_B = - \int_{\mathbb{C}_{\epsilon}}d^2z\,\vac{
\jmath^{(0)}_{\tb{m}}(u)\otimes
\bar\jmath^{(0)}_{\widebar{\tb{m}}}(\bar{u})\otimes
\jmath^{(0)}_{\tb{n}}(v)\otimes
\bar\jmath^{(0)}_{\widebar{\tb{n}}}(\bar{v})\,
\Omega_4(z,\bar{z})}_0\ .
\eeqa
Compared to the original $I_B$ we just dropped the group theoretic
factor that is associated with the zero modes. It may be reinstalled
easily.

While an integral similar to $\tilde I_B$ also appears for symmetric spaces
and was computed for these in \cite{Candu:2013cga}, we now have to pay
attention to the grading and the coefficients, in particular the nontrivial
$q_\A$, in the action. As a result, while a subset of terms turns out to
reproduce those found for symmetric spaces, we shall also find new
contributions that have no counterpart in the previous computation. To
begin with, we can rewrite the quantity \eqref{eq:CorrIB} in the form,
\begin{align}\notag
&
\tilde I_B = - \Pi\cdot\Bigg[
\sum_{\rho,\sigma=1}^r\sum_{\bar\rho,\bar\sigma=1}^{\bar r}
\Big\langle\,\jmath^{(0)}_{\tb{m}_\rho}(u)\otimes
\bar\jmath^{(0)}_{\widebar{\tb{m}}_{\bar\sigma}}(\bar{u})\,\otimes\, \jmath^{(0)}_{\tb{n}_\sigma}(v)\otimes \bar\jmath^{(0)}_{\widebar{\tb{n}}_{\bar\sigma}}(\bar{v})\,\Big\rangle_0
\otimes\\[2mm]
&\otimes
\int_{\mathbb{C}_\epsilon}\frac{d^2z}{\pi}
\vac{
\partial^{m_\rho}\phi(u)\otimes
\bar\partial^{\bar{m}_{\bar\rho}}\phi(\bar{u})\otimes
\partial^{n_{\sigma}}\phi(v)\otimes\bar\partial^{\bar{n}_{\bar\sigma}}
\phi(\bar{v})\,\Omega_4(z,\bar{z})}_0 \Bigg]\ .
\label{eq:sumdecomposition}
\end{align}
Here, $\jmath^{(0)}_{\tb{m}_\rho}$ denotes the tensor product
\eqref{eq:defproductcurrents2} of currents with the $\rho$-th factor
removed and we introduced a permutation $\Pi$ that acts on a tensor
power of $\mathfrak{m}$, see \cite{Candu:2013cga} for details. In
evaluating $\tilde I_B$ we insert the representation \eqref{eq:Omega4}
for the four vertex $\Omega_4$.

Let us begin with a general statement and consider the integrand of
eq. \eqref{eq:sumdecomposition} with some definite choice of $\mathbb{Z}_4$
grading for the currents,
$$
\I:=\vac{\partial^m\phi_\A(u)\!\otimes\!
\bar{\partial}^{\bar{m}}\phi_\B(\bar{u})\!\otimes\! \partial^n\phi_\C(v)\!\otimes\! \bar{\partial}^{\bar{n}}\phi_\D(\bar{v})\form{[\phi_\E,\partial\phi_\F]}
{[\phi_\G,\bar\partial\phi_\H]} (z,\bar{z})}_0.
$$
We can now split the symbols $\phi_\alpha= t_a \phi_\alpha^a$ into products
of fields and generators. Moving all generators to the left of the fields
and outside the correlator, we obtain
\begin{multline}
\I=(-1)^{|a|+|a||b|+|b|+|c||d|}(-1)^{|h|+|g||h|+|g|+|e||f|} t_a\otimes t_b\otimes t_c\otimes t_d\\\times
\Big\langle \partial^m\phi_\A^a(u)
\bar{\partial}^{\bar{m}}\phi_\B^b(\bar{u})\partial^n\phi_\C^c(v) \bar{\partial}^{\bar{n}}\phi_\D^d(\bar{v})(\phi_\E^e\partial\phi_\F^f\phi_\G^g\bar\partial\phi_\H^h)(z,\bar{z})f_{ef}^{\phantom{ef}i} f_{ghi}\Big\rangle_0\ .
\end{multline}
We observe that the sign factors we picked up while moving the generators
past fields cancel in all terms that are allowed by Wick contractions. After
inserting the  resulting expression for the integrand $\I$ back into the
$\tilde I_B$ we can perform the integration with the help of the
following integral formula that we derive in App.~\ref{sec:integralidentities}
\begin{multline}
\label{eq:integralid}
\int_{\mathbb{C}_{\epsilon}}\frac{d^2z}{\pi}\frac{a! b! c! d!}
{(z-u)^{a+1}(z-v)^{b+1}(\bar{z}-\bar{u})^{c+1}(\bar{z}-\bar{v})^{d+1}}
=\\=2\log\left|\frac{u-v}{\epsilon}\right|^2\times \frac{(-1)^{a+c}(a+b)!(c+d)!}
{(u-v)^{a+b+1}(\bar{u}-\bar{v})^{c+d+1}}+\text{non-log.}
\end{multline}
If we take all possible Wick contraction schemes into account and
make repeated use of the properties \eqref{eq:KY1loopconformal} of
our parameters we obtain
\begin{multline}
\tilde I_B= \Pi\cdot\Bigg[\sum_{\rho,\sigma=1}^r\sum_{\bar\rho,\bar\sigma=1}^{\bar r}\sum_{\A+\B+\C+\D=0} \Big \langle\,\jmath^{(0)}_{\tb{m}_\rho}(u)\otimes
\bar\jmath^{(0)}_{\widebar{\tb{m}}_{\bar\sigma}}(\bar{u})\,\otimes\, \jmath^{(0)}_{\tb{n}_\sigma}(v)\otimes \bar\jmath^{(0)}_{\widebar{\tb{n}}_{\bar\sigma}}(\bar{v})\,\Big\rangle_0\\
 \otimes\log\left|\frac{u-v}{\epsilon}\right|^2\,f_{m_\rho\bar m_{\bar\rho}n_\sigma\bar n_{\bar\sigma}}(u,v)
 t_a\otimes t_b\otimes t_c\otimes t_d\\
\times\left(-\left(-\frac43  + p_{\A+\C}-\frac 13(q_\A-q_\B+q_\C-q_\D)+q_{\A+\C}\right)(-1)^{|c|+|d|+|c||d|}H^{abcd}_{\A\B\C\D}\right.\\
+\left(-\frac23  +\frac{p_{\A+\D}}2-\frac 16(q_\A-q_\B+q_\C-q_\D)\right)(-1)^{|b|+|d|}A^{abcd}_{\A\B\C\D}\\
+\left.\left(-\frac23  +\frac{p_{\A+\B}}2-\frac 16(q_\A-q_\B+q_\C-q_\D)\right)(-1)^{|c|+|d|}X^{abcd}_{\A\B\C\D}\right)\Big]
+\text{non-log.} \nonumber
\end{multline}
where we introduced the following convenient shorthand that will appear frequently below,
\beq
f_{m\bar mn\bar n}(u,v)=\frac{(-1)^{n+m}(m+n-1)!(\bar m+\bar n-1)!}{R^6(u-v)^{m+n}(\bar{u}-\bar{v})^{\bar m+\bar n}}\,.
\eeq
In addition we have used the matrices $H,A$ and $X$ that were defined in
eqs.\ \eqref{eq:X}-\eqref{eq:H}. It follows from the Jacobi identity of
the Lie superalgebra $\mathfrak{g}$ that these matrices obey,
\beq
-(-1)^{|c|+|d|+|c||d|}H^{abcd}_{\A\B\C\D}=(-1)^{|b|+|d|}A^{abcd}_{\A\B\C\D}+
(-1)^{|c|+|d|}X^{abcd}_{\A\B\C\D}
\eeq
so that we are able to bring our result for $\tilde I_B$ into the
following form
\begin{multline}
\tilde I_B= \Pi\cdot\Bigg[\sum_{\rho,\sigma=1}^r\sum_{\bar\rho,\bar\sigma=1}^{\bar r}\sum_{\A+\B+\C+\D=0} \Big \langle\,\jmath^{(0)}_{\tb{m}_\rho}(u)\otimes
\bar\jmath^{(0)}_{\widebar{\tb{m}}_{\bar\sigma}}(\bar{u})\,\otimes\, \jmath^{(0)}_{\tb{n}_\sigma}(v)\otimes \bar\jmath^{(0)}_{\widebar{\tb{n}}_{\bar\sigma}}(\bar{v})\,\Big\rangle_0\\
 \otimes\log\left|\frac{u-v}{\epsilon}\right|^2\,f_{m_\rho\bar m_{\bar\rho}n_\sigma\bar n_{\bar\sigma}}(u,v)
 t_a\otimes t_b\otimes t_c\otimes t_d\\
\times\left(\left(2  - p_{\A+\C}+\frac 12(Q_{\A\C}-Q_{\B\D})\right)(-1)^{|c| + |d|+|c||d|}H^{abcd}_{\A\B\C\D}\right.\\\left.
+\frac{p_{\A+\D}}2 (-1)^{|b|+|d|} A^{abcd}_{\A\B\C\D}
+\frac{p_{\A+\B}}2 (-1)^{|c|+|d|} X^{abcd}_{\A\B\C\D}\right)\Bigg]+\text{non-log.}
\end{multline}
In order to read off the contributions to the dilation operator we still
have to reexpress the function $f(u,v)$ through currents. This is possible
and in the process we actually absorb all the sign factors that appear in
front of the matrices $H,A$ and $X$. As an example, let us consider the
terms involving $H$ and observe that
\begin{multline}\label{eq:4phicorr}
\vac{\partial^m\phi_\A(u)\otimes \bar{\partial}^{\bar{m}}\phi_\B(\bar{u})\otimes
\partial^n\phi_\C(v)\otimes \bar{\partial}^{\bar{n}}\phi_\D(\bar{v})}_0=
\\[2mm] =(-1)^{|a||b|+|a|+|b|} f_{m\bar m n \bar n}(u,v)
\ t_a\otimes t_b\otimes t^{a^{\prime}}\otimes
t^{b^{\prime}}\frac{\delta_{c,a^{\prime}}
\delta_{d,b^{\prime}}}{p_\A p_\B} \ .
\end{multline}
once this is inserted into our expression for $\tilde I_B$, we can read
off the corresponding contribution to the dilation operator from the
coefficient of the tree level correlation function. For details we
refer to section 3.2.1 of \cite{Candu:2013cga}. A similar analysis
may be performed for the terms involving $A$ and $H$. In these cases,
the relevant correlation functions are
$$\vac{\partial^m\phi_\A(u)\otimes \bar{\partial}^{\bar{m}}\phi_\B(\bar{u})
\otimes \bar\partial^{\bar n}\phi_\C(\bar v)\otimes {\partial}^{{n}}\phi_\D({v})}_0$$
and
$$\vac{\partial^m\phi_\A(u)\otimes {\partial}^{{n}}\phi_\C({v})\otimes \bar{\partial}^{\bar{m}}\phi_\B(\bar{u})\otimes
\bar\partial^{\bar n}\phi_\D(\bar v)}_0\ . $$
We leave all details to the reader. The general conclusion is that the factor
$f(u,v)$ along with the grading signs in front of $H,A$ and $X$ are absorbed.
Similar comments apply at many places below even if we refrain from stressing this
again.

\subsubsection{Contributions from case C}

Next we need to compute the part of the one-loop correction that arises from the
expansion of the currents. In our discussion we shall work with the component fields
$\phi_{i}:=(\phi,t_{i})$. When written in terms of these component fields, the
subleading part of the current becomes
$$\jmath^{(1)}=\half f^{ij}_{\phantom{ij}a}\phi_{i}\partial\phi_{j}t^{a}\ . $$
Here $i,j$ and $a$
run over a basis of the quotient space $\fm = \fg/\fh$.
In this case we can divide the discussion in diagonal and off-diagonal
contributions. We call diagonal those contributions that do not cancel
at tree level, i.e. come from correlation functions of two fields that
have complementary grading of the (anti)holomorphic tails.

\subsubsection*{Diagonal contributions}
In this case only one type of terms appears which looks as follows
\begin{equation}
	\begin{split}
		\langle \partial^{m}\jmath^{(1)}(u)&\otimes\partial^{n}\jmath^{(1)}(v)\rangle =
		\\[2mm]
		= -\frac{(-1)^{m+\grade{a}}}{4 R^{4}}
		&\ln\abs{\frac{u-v}{\varepsilon}}^2 \frac{\fact{(m+n+1)}}{(u-v)^{m+n+2}}
		f^{ij}_{\phantom{ij}a}
		f^{kl}_{\phantom{kl}b} \eta_{ik}\eta_{jl} t^{a}\otimes
		t^{b}+\orderepsilon.
	\end{split}
	\label{eqn:tail_one_loop}
\end{equation}
Similar contributions arise from the anti-holomorphic currents.
If we take into account that the Killing form of $G$ vanishes by
assumption and then compare with eq.\ (5.8) in \cite{Berkovits:1999zq}
we can identify the combination of the structure constants that appears
in the previous expression with the Ricci tensor of the coset space,
i.e.\
\begin{equation}
	f^{ij}_{\phantom{ij}a}
	f^{kl}_{\phantom{kl}b} \eta_{ik}\eta_{jl} = 4R_{ab}(G/H) \ .
	\label{eqn:tail_one_loop_ricci}
\end{equation}
Below we shall see that a similar term involving the Ricci tensor also
arises from case E. The latter actually cancels the contributions from
case C.

\subsubsection*{Off-diagonal contributions}
We take an example. From case C
we have two possible contributions, that contribute equally, namely
$$
X = \frac{1}{4}\langle\partial^m\phi_3(u)
\otimes \com{\phi_2}{\bar\partial^{\bar  m} \phi_1}(u,\bar u)\otimes \com{\phi_2}{\partial^n\phi_1}(v,\bar{v})\otimes
\bar{\partial}^{\bar n}\phi_3(\bar{v})\rangle
$$
and
$$ \tilde X =
\frac{1}{4} \langle\com{\phi_2}{\partial^m\phi_1}(u,\bar u)
\otimes \bar{\partial}^{\bar m}\phi_3(\bar u)\otimes\partial^n\phi_3 (v)\otimes
\com{\phi_2}{\bar\partial^{\bar n}\phi_1}(v,\bar{v})\rangle\ .
$$
In both expressions we have moved all the  derivatives into the
commutator, using the fact that terms in which derivatives act on
the non-derivative field cannot give rise to logarithms. The sum
of the previous two quantities is easily seen to give
\begin{align}\label{eq:j'barj'}
& X + \tilde X = -\frac 12\log\left|\frac{u-v}{\epsilon}\right|^2
 t_{3i}\otimes t_{3j}\otimes t_{3k}\otimes t_{3l} \form{\com{t_1^l}{t_1^j}}{\com{t_{1}^k}{t_{1}^i}}\times
 \nonumber \\ & \hspace*{2cm} \,\times f_{m\bar mn\bar n}(u,v)+\text{non-log.}\
\end{align}
The results of this analysis can be summarized through the following
expression for the off-diagonal contributions to $\tilde I_C$,
\begin{multline}
\tilde I_C^{\text{off}}= \Pi\cdot\Bigg[\sum_{\rho,\sigma=1}^r\sum_{\bar\rho,\bar\sigma=1}^{\bar r}\sum_{\A+\B+\C+\D=0} \Big \langle\,\jmath^{(0)}_{\tb{m}_\rho}(u)\otimes
\bar\jmath^{(0)}_{\widebar{\tb{m}}_{\bar\sigma}}(\bar{u})\,\otimes\, \jmath^{(0)}_{\tb{n}_\sigma}(v)\otimes \bar\jmath^{(0)}_{\widebar{\tb{n}}_{\bar\sigma}}(\bar{v})\,\Big\rangle_0\\
\otimes\frac{p_{\A+\C}}2\log\left|\frac{u-v}{\epsilon}\right|^2 f_{m_\rho\bar m_{\bar\rho}n_\sigma\bar n_{\bar\sigma}}(u,v)\\
\,t_a\otimes t_b\otimes t_c\otimes t_d
 (-1)^{|c|+|d|+ |c||d|}H^{abcd}_{\A\B\C\D}\bigg]
\end{multline}
where we introduced $p_{\A+\C}$ to implement the fact that these terms
are off-diagonal.

\subsubsection{Contributions from case D}

In the calculation we shall make use of
the following integral formula
\begin{equation}
	\begin{split}
		\int_{\mathbb{C}_{\varepsilon}}^{}&\frac{\dd^{2} z}{\pi}
		\frac{1}{(z-x)^{a+1}(\bar{z}-\bar{x})^{b+1}(\bar{z}-\bar{y})^{c+1}}
		=
		\\
		&=\delta_{a,0}\frac{(-1)^{b+1}}{a!}\binom{b+c}{c}
		 \frac{\ln\left\lvert\frac{x-y}{\varepsilon}\right\rvert^{2}}{(\bar{x}-\bar{y})^{b+c+1}}
		+\mathcal{O}(\varepsilon)
	\end{split}
	\label{eqn:integral2}
\end{equation}
This formula is derived in App.\ \ref{sec:integralidentities}. Let us point
out that logarithmic singularities only exist when $a =0$. This
implies that most
terms that appear in the evaluation of case D do not contain any
logarithmic divergencies. Having made this observation, let us
discuss the contributions from $I_D$. Once again we shall
distinguish between diagonal and off-diagonal contributions.

\subsubsection*{Diagonal contribution}
Recall that $\Omega_3 = \Omega_3^a$ and that we can add a
total derivative in order to bring $\Omega_3^a$ into the simple form
$\Omega_3^{\prime a}$. This is the form we shall use.
We address the
two terms of $\Omega_3^{\prime a}$ separately. Given our introductory
comment, there are few terms that contribute. The following integral
is an example
\begin{equation}
	\begin{split}
		\int_{\mathbb{C}_{\varepsilon}}^{}\frac{\dd^{2} z}{\pi}
		\langle \com{\phi_3}{\partial^{m+1}\phi_2}(u,\bar{u})
		\otimes\partial^{n+1}\phi_3(v)
		(\partial\phi_{1},[\phi_{2},\bar{\partial}\phi_{1}])(z,\bar{z})
		\rangle =\\
		 \frac{(-1)^{m+1}}{R^{4}}
		\ln\abs{\frac{u-v}{\varepsilon}}^2
		\frac{\fact{(m+n+1)}}{(u-v)^{m+n+2}} \com{t_{2i}}{t_{3j}}\otimes
		\com{t_2^i}{t_1^{j}} + \text{non-log.}
	\end{split}
	\label{eqn:one_loop_tail_three_vertex_scetch}
\end{equation}
This term is actually canceled by the following contact term
\beqa
		\int_{\mathbb{C}_{\varepsilon}}^{}\frac{\dd^{2} z}{\pi}
		\langle \com{\phi_2}{\partial^{m+1}\phi_3}(u,\bar{u})
		\otimes\partial^{n+1}\phi_3(v)
		(\partial\phi_{1},[\phi_{2},\bar{\partial}\phi_{1}])(z,\bar{z})\nonumber
		\rangle =\\
-\frac{1}{R^{4}}\partial_u^n\partial_v^m\int_{\mathbb{C}_{\varepsilon}}^{}\frac{\dd^{2} z}{\pi}
		\ln\abs{\frac{u-z}{\varepsilon}}^2\frac 1{u-z}\delta^2(v-z)
		 \com{t_{2i}}{t_{3j}}\otimes
		\com{t_2^i}{t_1^{j}} =\\
		-\frac{(-1)^{m+1}}{R^{4}}
		\ln\abs{\frac{u-v}{\varepsilon}}^2
		\frac{\fact{(m+n+1)}}{(u-v)^{m+n+2}} \com{t_{2i}}{t_{3j}}\otimes
		\com{t_2^i}{t_1^{j}} + \text{non-log.}\nonumber
\eeqa
A similar cancelation occurs if we consider terms in which the current at $v$
is expanded and the second contribution of $\Omega_3^{\prime a}$ employed in
the contractions. Following this reasoning one may see that all diagonal
contributions for case D cancel each other.

\subsubsection*{Off-diagonal contribution}

In this case we shall work with the generic expression \eqref{eq:Omega3} for
$\Omega_3$. First we consider the expansion of one of the holomorphic currents
\begin{equation}
    \begin{split}
        \int_{\mathbb{C}_{\varepsilon}}^{}\frac{\dd^{2} z}{\pi}
        \langle (\com{\phi_\E}{\partial^{m+1}\phi_\F}\otimes\bar
\partial^{\bar m+1} \phi_\B)(u,\bar{u})
        \otimes(\partial^{n+1}\phi_\C\otimes\bar\partial^{\bar n+1}
\phi_\D)(v,\bar v)\\
\times(\partial\phi_\G,[\phi_\H,\bar{\partial}\phi_\iota])(z,\bar{z})
        \rangle \ .
    \end{split}
\end{equation}
from this we get
\beqa
-\frac14 Q_{\B\D}\log\left|\frac{u-v}{\epsilon}\right|^2 f_{m\bar mn\bar n}(u,v)
\,t_a\otimes t_b\otimes t_c\otimes t_d(-1)^{|c|+|d|+
|c||d|}H^{abcd}_{\A\B\C\D}\ ,
\eeqa
and similarly for the expansion of the other holomorphic current. If we consider
the expansion of the anti-holomorphic current we obtain instead
\beqa
+\frac 12 Q_{\A\C}\log\left|\frac{u-v}{\epsilon}\right|^2 f_{m\bar mn\bar n}(u,v)
\,t_a\otimes t_b\otimes t_c\otimes t_d(-1)^{|c|+|d|+ |c||d|} H^{abcd}_{\A\B\C\D}\ .
\eeqa
Summing all the contributions we find
\begin{multline}
\tilde I_D^{\text{off}}= \frac12\Pi\cdot\Bigg[\sum_{\rho,\sigma=1}^r
\sum_{\bar\rho,\bar\sigma=1}^{\bar r}\sum_{\A+\B+\C+\D=0} \Big\langle\,\jmath^{(0)}_{\tb{m}_\rho}(u)\otimes
\bar\jmath^{(0)}_{\widebar{\tb{m}}_{\bar\sigma}}(\bar{u})\,\otimes\, \jmath^{(0)}_{\tb{n}_\sigma}(v)\otimes
\bar\jmath^{(0)}_{\widebar{\tb{n}}_{\bar\sigma}}(\bar{v})\,\Big\rangle_0\\
(Q_{\A\C}-Q_{\B\D})\log\left|\frac{u-v}{\epsilon}\right|^2 f_{m_\rho\bar m_{\bar\rho}n_\sigma\bar n_{\bar\sigma}}(u,v)\\
\,t_a\otimes t_b\otimes t_c\otimes t_d (-1)^{|c|+|d|+|c||d|}
H^{abcd}_{\A\B\C\D}\Bigg]\ .
\end{multline}
This contribution is antisymmetric in fields of grading 1 and 3 and it is
identical to the antisymmetric contribution from case B.

\subsubsection{Contributions from case E}

Let us now address the final case with two insertions of the vertex
$\Omega_3$. Recall that in a theory with vanishing one-loop
beta-function, the three vertex is purely antisymmetric, i.e.\
$\Omega_3= \Omega_3^a$. To ease the calculation we added a total
derivative. Throughout this section we shall use the three vertex
$\Omega_3^{\prime a}$, as we did before.

In the calculation we have two types of terms. The first type consists
of  contributions which involve two contractions between the two $\Omega_3$
and two more contractions between the vertices and two currents. These only contribute to the diagonal part. On the
other hand, we can also have terms in which there is a single contraction
between the two vertices and four contractions with the tails of currents.
We shall argue that the first type cancels the contributions from case C
while the second type contributes new terms to the dilation operator.

In the case of two contractions we get only terms that contribute to the diagonal
part of the anomalous dimension.

\beqa\nonumber
	& & \hspace*{-1cm} \frac 18\int_{\mathbb{C}_{\varepsilon}}^{}\frac{\dd^{2} z}{\pi}
\int_{\mathbb{C}_{\varepsilon}}^{}\frac{\dd^{2} w}{\pi}
		\big< {\partial^{m+1}\phi_2}(u)
		\otimes\partial^{n+1}\phi_2(v) (\partial\phi_{1},[\phi_{2},\bar{\partial}\phi_{1}])
(z,\bar{z})\times \\
& & \hspace*{5cm} \times (\partial\phi_{3},[\phi_{2},\bar{\partial}\phi_{3}])(w,\bar{w})
		\big>_0 =
		\\[2mm]
		& & = -\frac{(-1)^{m}}{8 R^{4}}
		\ln\abs{\frac{u-v}{\varepsilon}}^2 \frac{\fact{(m+n+1)}}{(u-v)^{m+n+2}}
		f^{ij}_{\phantom{ij}a}
		f^{kl}_{\phantom{kl}b} \eta_{ik}\eta_{jl} t_2^{a}\otimes
		t_2^{b}+ \text{non-log.}\nonumber
\eeqa
Here we have inserted the first term of $\Omega_3^{\prime a}$ at $(z,\bar z)$
and the second term at $(w,\bar w)$. The opposite choice turns out to
give exactly the same result so that we get a numerical prefactor -$1/4$
in place of -$1/8$ after summing both contributions. We see that the
result cancels against the contribution from case C for $a$ and $b$
labeling basis elements of $\fg_2$,  i.e.\ when $|a|= |b| = 0$. In
evaluating the relevant integrals, we have used the formula
\beqa \nonumber
& & \int_{\mathbb{C}_{\varepsilon}}^{}\frac{\dd^{2} z}{\pi}\int_{\mathbb{C}_{\varepsilon}}^{}\frac{\dd^{2} z}{\pi}
		\frac{1}{(z-x)({w}-{y})(z-w)^2(\bar{z}-\bar{w})^{2}}
		= \\[2mm] & = &  -\int_{\mathbb{C}_{\varepsilon}}^{}\frac{\dd^{2} z}{\pi}\frac{1}{(z-x)}\frac{1}{(z-y)^2}\frac{1}{(\bar z-\bar y)}
		=\frac{\ln\abs{\frac{u-v}{\varepsilon}}^2}{(x-y)^2(\bar x-\bar y)^2}
\nonumber
\eeqa
and derivatives thereof. This can be derived using formulas in
Appendix \ref{sec:integralidentities}.
We can perform a similar analysis in case the currents from the tails possess
odd grade rather than even. The result is
\beqa\nonumber
	& & \hspace*{-1cm} \frac 18\int_{\mathbb{C}_{\varepsilon}}^{}\frac{\dd^{2} z}{\pi}\int_{\mathbb{C}_{\varepsilon}}^{}\frac{\dd^{2} z}{\pi}
		\big< {\partial^{m+1}\phi_1}(u)
		\otimes\partial^{n+1}\phi_3(v) (\partial\phi_{1},[\phi_{2},\bar{\partial}\phi_{1}])(z,\bar{z})\times \\[2mm]
& & \hspace*{5cm} \times (\partial\phi_{3},[\phi_{2},\bar{\partial}\phi_{3}])(w,\bar{w})
		\big>_0 =
		\\[2mm]
		& & = \frac{(-1)^{m}}{8 R^{4}}
		\ln\abs{\frac{u-v}{\varepsilon}}^2 \frac{\fact{(m+n+1)}}{(u-v)^{m+n+2}}
		f^{ij}_{\phantom{ij}a}
		f^{kl}_{\phantom{kl}b} \eta_{ik}\eta_{jl} t_1^{a}\otimes
		t_3^{b}+ \text{non-log.}\nonumber
\eeqa
In the process we have used the integral formula \eqref{eqn:logintegral_2}
from Appendix \ref{sec:integralidentities}. Once again, the result
cancels the contribution from case C.

Let us then turn to the second type of terms in which we have four
contractions between the vertices and the tail of currents. As above,
we illustrate the computations with a concrete example,
\beqa
&&  \hspace*{-5mm}\frac18\int_{\mathbb{C}_{\epsilon}}\frac{\dd^2z}{\pi}\int_{\mathbb{C}_{\varepsilon}}^{}\frac{\dd^{2} z}{\pi}\langle\partial^{m}\phi_1(u)\otimes
\bar \partial^{\bar{m}}\phi_2(\bar u)\otimes\partial^{n}\phi_{3}(v)\otimes
\bar{\partial}^{\bar n}\phi_{2}(\bar{v})\,\times\nonumber\\[2mm]
&&\hspace*{2cm} \big(\form{\partial\phi_1}{\com{\phi_2}{\bar\partial\phi_1}}(z,\bar z)
\form{\partial\phi_{3}}{\com{\phi_{2}}{\bar{\partial}\phi_{3}}}(w,\bar{w})\rangle_{0}=
\\[2mm]
&& \nonumber=\frac14\ln\abs{\frac{u-v}{\varepsilon}}^2 f_{m\bar mn\bar n}(u,v)\times\notag\\[2mm]
  &&\hspace*{-8mm} \nonumber t_{1i}\otimes t_{2j}\otimes t_{3k}\otimes t_{2l} \big[\form{\com{t_2^l}{t_1^k}}{\com{t_{3}^i}{t_{2}^j}}-
 \form{\com{t_2^j}{t_1^k}}{\com{t_{3}^i}{t_{2}^l}}\big]+ \text{non-log.}
\eeqa
If we insert the second non-trivial term in $\Omega_3^{\prime a}$ we obtain
an identical contribution so that we just have to multiply the right hand
side of the previous formula by a factor of two.

There is another qualitatively somewhat different example we want
to consider in which the only contraction between the two vertices is a
contraction between non-derivative fields. This happens e.g. in
\beqa
&&  \hspace*{-1cm}\frac18\int_{\mathbb{C}_{\epsilon}}\frac{\dd^2z}{\pi}\int_{\mathbb{C}_{\varepsilon}}^{}\frac{\dd^{2} z}{\pi}\langle\partial^{m}\phi_1(u)\otimes
\bar \partial^{\bar{m}}\phi_1(\bar u)\otimes\partial^{n}\phi_{3}(v)\otimes
\bar{\partial}^{\bar n}\phi_{3}(\bar{v})\,\times          \nonumber\\[2mm]
&&\hspace*{1cm} \big(\form{\partial\phi_1}{\com{\phi_2}{\bar\partial\phi_1}}(z,\bar z)
\form{\partial\phi_{3}}{\com{\phi_{2}}{\bar{\partial}\phi_{3}}}(w,\bar{w})\rangle_{0}=
\\[2mm]
&& \hspace*{-5mm}
\nonumber =-\frac14\ln\abs{\frac{u-v}{\varepsilon}}^2 f_{m\bar mn\bar n}(u,v)\times
\notag\\[2mm]
  &&\hspace*{5mm}
 t_{1i}\otimes t_{1j}\otimes t_{3k}\otimes t_{3l} \form{\com{t_1^l}{t_1^k}}{\com{t_{3}^i}{t_{3}^j}}+ \text{non-log.}\nonumber
\eeqa
This and similar terms can be evaluated using the integral formula
\eqref{eq:logintegral_1a}. Off diagonal terms are evaluated similarly and
all the relevant integrals can be found in appendix A. The final result
reads
\begin{multline}
\tilde I_E= \Pi\cdot\Bigg[\sum_{\rho,\sigma=1}^r\sum_{\bar\rho,\bar\sigma=1}^{\bar r}\sum_{\A+\B+\C+\D=0} \Big\langle\,\jmath^{(0)}_{\tb{m}_\rho}(u)\otimes
\bar\jmath^{(0)}_{\widebar{\tb{m}}_{\bar\sigma}}(\bar{u})\,\otimes\, \jmath^{(0)}_{\tb{n}_\sigma}(v)\otimes \bar\jmath^{(0)}_{\widebar{\tb{n}}_{\bar\sigma}}(\bar{v})\,\Big\rangle_0\\
\otimes\frac{(-1)^{|a|+|c|}}2\log\left|\frac{u-v}{\epsilon}\right|^2 f_{m_\rho\bar m_{\bar\rho}n_\sigma\bar n_{\bar\sigma}}(u,v)\,
t_a\otimes t_b\otimes t_c\otimes t_d \\
 \times\left(- Q_{\A\C}Q_{\B\D}(-1)^{|c|+|d|+|c||d|}H^{abcd}_{\A\B\C\D}
+Q_{\A\D}Q_{\B\C} (-1)^{|b|+|d|}A^{abcd}_{\A\B\C\D}\right.\\\left.
+Q_{\A\B}Q_{\C\D} (-1)^{|c|+|d|}X^{abcd}_{\A\B\C\D}\right)\Bigg]\ .
\end{multline}

\subsubsection{Contributions from case F and G}
These two cases turn out to contribute only off-diagonal terms.
Contributions from case G are symmetric in the exchange of the
grading 1 and 3, those from case F are antisymmetric. Note that
these contributions appear only if all the fields in the tail
of the two fields have complementary grading except for one.
This means that from these terms we read the off-diagonal entries
of the dilation operator that, at one loop, change the grading
of one of the fields in the tail at the time. In case G the
evaluation of the new terms is fairly straightforward since no
integrals need to be performed. The calculation of F is very
similar to the one for case D. Without presenting any more
details we state the answer for two vertex operators that
differ in the grading of one holomorphic field,
\begin{multline}
I_F+I_G=-\Pi\cdot\int_{G/H}d\mu(g_0H)\,\Bigg[\sum_{\rho,\sigma=1}^r\sum_{\A,\B=1,2,3}
V^{(1)}(t_c)\otimes V^{(0)}\\\otimes\jmath^{(0)}_{\tb{m}_\rho}\otimes t_a\otimes \jmath^{(0)}_{\widebar{\tb{m}}} \otimes\jmath^{(0)}_{\tb{n}_\sigma}\otimes t_b\otimes \jmath^{(0)}_{\widebar{\tb{n}}}\\
\times(1+(-1)^{|c|}Q_{\A\B})\\
\frac{(-1)^{m_\rho}}{4 R^{4}}
		\ln\abs{\frac{u-v}{\varepsilon}}^2
\frac{\fact{(m_\rho+n_\sigma+1)}}{(u-v)^{m_\rho+n_\sigma+2}}f^{cab}\Bigg]\ .
\end{multline}
A similar results applies for the anti-holomorphic case. This concludes our
derivation of the one-loop dilation operator.

\section{Conclusion}
In this paper we calculated the one-loop dilation operator on the
space of all vertex operators in NLSMs on semi-symmetric coset
superspaces, at least for a special choice of the background
parameters that appears in pure spinor and hybrid formulations
and makes the one-loop beta-function vanish. While the diagonal
terms are identical to the ones found for conformal NLSMs on
symmetric spaces \cite{Candu:2013cga}, there exist interesting new
off-diagonal contributions to the dilation operators. These
involve a new XXZ-like interaction term between left and right
moving currents in the vertex operator. Of course, it would be
very interesting to diagonalize the dilation operator for
general states, not just for the small subsector of fields we
considered in section \ref{sec:summary}.2.

In the case of symmetric spaces it was very easy to extend the analysis beyond
the conformal case. In fact, \cite{Candu:2013cga} discusses a simple additional
term that must be added to the formula for anomalous dimensions in order to be
applicable for symmetric spaces $G/H$ with numerator group $G$ other than
$G = PSU(N|N), OSP(2N+2|2N)$ or $D(2,1;\alpha)$ when the corresponding sigma
model ceases to be conformal. The extension of our analysis for semi-symmetric
spaces to couplings with non-vanishing beta-function, however, would certainly
require significantly more work and not lead to a simple result.

Let us stress once again that the calculations we have performed apply to all
NLSMs that appear in the hybrid formulation of $AdS$ backgrounds\cite{Berkovits:1999zq,Berkovits:1999im}. This includes
e.g.\ the coset $PSU(1,1|2)/U(1)\times U(1)$ whose bosonic base is $AdS_2\times
S^2$ \cite{Berkovits:1999zq}. It would be interesting to list all vertex operators for this case,
to compute the full one-loop partition function and to extend the required
input from harmonic analysis to higher dimensional $AdS$ backgrounds for
which the stability subgroup $H$ is non-compact.

A key motivation for this work arises from the challenge to find a dual
world-sheet description of strongly curved $AdS$ backgrounds. In order
to test any future proposal for such a 2-dimensional dual model, one
will have to compare at least parts of the spectra. We have recently
outlined the contours of such investigations at the example of the
supersphere sigma model \cite{Cagnazzo:2014yha}. The target space
$S^{3|2}$ that was used in the analysis is a symmetric superspace.
In this case, a dual Gross-Neveu model description had been proposed
by Candu and Saleur. We showed how recent all-loop results in deformed
WZNW models could be used to test this duality by matching the spectrum
of the free sigma model and its one-loop corrections. The results of the
present work, which can be straightforwardly applied to the sigma model
e.g. on $AdS_2 \times S^2$, provide plenty of data that could be matched
with a potential dual formulation. Unlike the supersphere model, for which
no string embedding is known, the $AdS_2 \times S^2$ model can be made into
a true string theory. This brings in additional techniques which could
complement the analytic tools developed here and help to resolve some of
the issues that were left open in our previous investigation of supersphere
models, such as e.g. the role of strongly RG relevant high gradient operators.

In this work we have not discussed the relation between various different
formulations of superstring theory and we focused on the models that appear
in the hybrid formalism. It would certainly be interesting to repeat our
analysis within the ($\kappa$-gauge fixed) Green-Schwarz formulation.

\appendix
\section{Derivation of integral formulas}
\label{sec:integralidentities}
First of all we recall a useful formula that was derived in \cite{Candu:2013cga}
\beq\label{eq:4integral}
\int_{\mathbb{C}_{\epsilon}}\frac{d^2z}{\pi}\frac{1}{(z-x)(z-y)(\bar{z}-\bar{x})(\bar{z}-\bar{y})}=\frac{2\log\left|\frac{x-y}{\epsilon}\right|^2}{|x-y|^2}+\calO(\epsilon)\ .
\eeq
By taking now the appropriate number of derivatives in $x$, $y$, $\bar{x}$ or $\bar{y}$ on both sides of the above equation, we recover eq.~\eqref{eq:integralid}.

Another useful integral is
\begin{equation}\label{eq:3integral}
	\begin{split}
		\int_{\mathbb{C}_{\varepsilon}}^{}&\frac{\dd^{2} z}{\pi}
		\frac{1}{(z-x)(\bar{z}-\bar{x})(\bar{z}-\bar{y})}
		=\frac{1}{\bar{x}-\bar{y}}\int_{\mathbb{C}_{\varepsilon}}^{}
		\frac{\dd^{2} z}{\pi}
		\bar{\partial}_{z}
		\frac{\ln\abs{\frac{z-x}{z-y}}^{2}}{z-x}
		\\
		&=\frac{-1}{\bar{x}-\bar{y}}\oint_{\partial\mathbb{C}_{\varepsilon}}
		\frac{\dd z}{2\pi i}
		\frac{\ln\abs{\frac{z-x}{z-y}}^{2}}{z-x}
		=
		-\frac{\ln\abs{\frac{x-y}{\varepsilon}}^{2}}{(\bar{x}-\bar{y})}
		+\mathcal{O}(\varepsilon)
	\end{split}
\end{equation}
Note the differences with the previous case with four factors in the
denominator. In particular, only taking derivatives with respect to
barred variables retains the logarithmic factor. This explains the
delta factor in the above formula. The difference by a factor
\half{} is due to the difference in the number of poles.

Using \eqref{eq:4integral} and \eqref{eq:3integral} and their derivatives we can calculate a
series of double integrals:

\begin{equation}
	\begin{split}
		\int_{\dom}\frac{d^{2}z}{\pi}
		\int_{\dom}\frac{d^{2}w}{\pi}
		&\frac{1}{u-z}\frac{1}{v-w}\frac{1}{(\bar{u}-\bar{z})^{2}}
		\frac{1}{(\bar{v}-\bar{w})^{2}}\frac{1}{(z-w)^{2}}=
		\\
		&=-\frac{2\ln\lvert\frac{u-v}{\varepsilon}\rvert^{2}}{(u-v)^{2}
		(\bar{u}-\bar{v})^{2}}
	\end{split}
	\label{eqn:integral_1}
\end{equation}
That can be derived by starting with the $w$-integral:
\begin{equation}
	\begin{split}
		&\qquad\int_{\dom}\frac{d^{2}w}{\pi}
		\frac{1}{v-w}\frac{1}{(\bar{v}-\bar{w})^{2}}\frac{1}{(z-w)^{2}}=
		\\
		&=-\bar{\partial}_{v}\partial_{z} \int_{\dom}\frac{d^{2}w}{\pi}
		\frac{1}{w-v}\frac{1}{\bar{w}-\bar{v}}\frac{1}{w-z}=
		\\
		&=\bar{\partial}_{v}\partial_{z}
		\frac{\ln\lvert\frac{v-z}{\varepsilon}\rvert^{2}}{v-z} + \oep
		=\frac{1}{(v-z)^{2}}\frac{1}{\bar{v}-\bar{z}}+\oep
	\end{split}
	\label{eqn:integral_1_der_1}
\end{equation}
Plugging this in the double integral:
\begin{equation}
	\begin{split}
		\int_{\dom}\frac{d^{2}z}{\pi}
		&\frac{1}{u-z}\frac{1}{(v-z)^{2}}\frac{1}{(\bar{u}-\bar{z})^{2}}
		\frac{1}{\bar{v}-\bar{z}}=
		\\
		&=-\frac{2\ln\lvert\frac{u-v}{\varepsilon}\rvert^{2}}{(u-v)^{2}
		(\bar{u}-\bar{v})^{2}}
	\end{split}
	\label{eqn:integral_1_der_2}
\end{equation}
Similarly one can calculate:
\begin{equation}
	\begin{split}
		\int_{\dom}\frac{d^{2}z}{\pi}
		\int_{\dom}\frac{d^{2}w}{\pi}
		&\frac{1}{u-w}\frac{1}{v-z}\frac{1}{(\bar{u}-\bar{z})^{2}}
		\frac{1}{(\bar{v}-\bar{w})^{2}}\frac{1}{(z-w)^{2}}=
		\\
		&=+\frac{2\ln\lvert\frac{u-v}{\varepsilon}\rvert^{2}}{(u-v)^{2}
		(\bar{u}-\bar{v})^{2}}
	\end{split}
	\label{eqn:integral_2}
\end{equation}

\begin{equation}
	\begin{split}
		\int_{\dom}\frac{d^{2}z}{\pi}
		\int_{\dom}\frac{d^{2}w}{\pi}
		&%
		\frac{1}{(u-z)^{2}}
		\frac{1}{(v-w)^{2}}
		\frac{1}{(\bar{u}-\bar{z})^{2}}
		\frac{1}{\bar{v}-\bar{w}}
		\frac{1}{z-w}
		=
		\\
		&=-\frac{2\ln\lvert\frac{u-v}{\varepsilon}\rvert^{2}}{(u-v)^{2}
		(\bar{u}-\bar{v})^{2}}
	\end{split}
	\label{eqn:integral_3}
\end{equation}

\begin{equation}
	\begin{split}
		\int_{\dom}\frac{d^{2}z}{\pi}
		\int_{\dom}\frac{d^{2}w}{\pi}
		&%
		\frac{1}{(v-z)^{2}}
		\frac{1}{(u-w)^{2}}
		\frac{1}{(\bar{u}-\bar{z})^{2}}
		\frac{1}{\bar{v}-\bar{w}}
		\frac{1}{z-w}
		=
		\\
		&=+\frac{2\ln\lvert\frac{u-v}{\varepsilon}\rvert^{2}}{(u-v)^{2}
		(\bar{u}-\bar{v})^{2}}
	\end{split}
	\label{eqn:integral_4}
\end{equation}

In the main text we need also some double integrals containing logarithms. For
this reason we need to calculate:
\begin{equation}
	\begin{split}
		\int_{\dom}\frac{d^{2}w}{\pi}
		&%
		\frac{1}{(u-w)^{2}}
		\frac{1}{(\bar u-\bar w)^{2}}
		\ln\abs{\frac{z-w}{\varepsilon}}^2
		=
		\\
		&=\partial_u\partial_{\bar u}\int_{\dom}\frac{d^{2}w}{\pi}\partial_{\bar w}\left(\frac{\ln\left(\frac{\bar u-\bar w}{\bar u-\bar z}\right) \ln\abs{\frac{z-w}{\varepsilon}}^2+\text{Li}_2 \left(\frac{\bar w-\bar z}{\bar u-\bar z}\right)}{w-u}\right)
		=
		\\
		&=\partial_u\partial_{\bar u}\left(\frac{\pi^2}{6}+\ln\left(\frac{\varepsilon}{\bar u-\bar z}\right) \ln\abs{\frac{u-z}{\varepsilon}}^2\right)=\
		\\
		&=-\frac{1}{u-z}\frac{1}{\bar u-\bar z}
		\end{split}
	\label{eqn:logintegral_1}
\end{equation}

Where we have used the fact that $\text{Li}_2(1)=\frac{\pi^2}{6}$.
This result can be used to calculate:
\begin{equation}
	\begin{split}
		\int_{\dom}\frac{d^{2}z}{\pi}
		\int_{\dom}\frac{d^{2}w}{\pi}
		&\frac{1}{(u-w)^2}\frac{1}{(v-z)^2}\frac{1}{(\bar{u}-\bar{w})^{2}}
		\frac{1}{(\bar{v}-\bar{z})^{2}}\ln\abs{\frac{z-w}{\varepsilon}}^2 =
		\\
		&=-\frac{2\ln\lvert\frac{u-v}{\varepsilon}\rvert^{2}}{(u-v)^{2}
		(\bar{u}-\bar{v})^{2}}
	\end{split}\label{eq:logintegral_1a}
\end{equation}

Other useful logarithmic integrals are:
\begin{equation}
	\begin{split}
		\int_{\dom}\frac{d^{2}w}{\pi}
		&%
		\frac{1}{(u-w)^{2}}
		\frac{1}{(\bar z-\bar w)^{2}}
		\ln\abs{\frac{z-w}{\varepsilon}}^2
		=
		\\
		&=-\partial_u\int_{\dom}\frac{d^{2}w}{\pi}\partial_{\bar w}\left(\frac{1+ \ln\abs{\frac{z-w}{\varepsilon}}^2}{(w-u)(\bar w-\bar z)}\right)
		=
		\\
		&=-\partial_u\left(\frac{1+ \ln\abs{\frac{z-u}{\varepsilon}}^2}{(\bar u-\bar z)}\right)=\
		\\
		&=-\frac{1}{u-z}\frac{1}{\bar u-\bar z}
		\end{split}
	\label{eqn:logintegral_2}
\end{equation}
and
\begin{equation}
	\begin{split}
		\int_{\dom}\frac{d^{2}w}{\pi}
		&%
		\frac{1}{(u-w)^{2}}
		\frac{1}{(\bar v-\bar w)^{2}}
		\ln\abs{\frac{z-w}{\varepsilon}}^2
		=
		\\
		&=\partial_u\partial_{\bar v}\int_{\dom}\frac{d^{2}w}{\pi}\partial_{\bar w}\left(\frac{\ln\left(\frac{\bar v-\bar w}{\bar v-\bar z}\right) \ln\abs{\frac{z-w}{\varepsilon}}^2+\text{Li}_2 \left(\frac{\bar w-\bar z}{\bar v-\bar z}\right)}{w-u}\right)
		=
		\\
		&=\partial_u\partial_{\bar u}\left(\text{Li}_2 \left(\frac{\bar u-\bar z}{\bar v-\bar z}\right)+\ln\left(\frac{\varepsilon}{\bar u-\bar z}\right) \ln\abs{\frac{u-z}{\varepsilon}}^2\right)=\
		\\
		&=\frac{(\bar z-\bar u)}{(\bar u-\bar v)(u-z)(\bar v-\bar z)}
		\end{split}
	\label{eqn:logintegral_3}
\end{equation}


\end{document}